\documentclass[reprint, superscriptaddress, secnumarabic, amssymb,aps,prl]{revtex4-2}

\usepackage{chemformula,siunitx}
\usepackage{lastpage}
\usepackage[protrusion=true, expansion=true]{microtype}
\usepackage{wrapfig}
\usepackage{multirow}
\usepackage{array}
\usepackage{ragged2e}
\usepackage{mathrsfs}
\justifying

\usepackage{epstopdf}
\usepackage[T1]{fontenc}
\usepackage{amsbsy}
\usepackage[T1]{fontenc}
\usepackage{amsmath}
\usepackage{amssymb}
\usepackage{bbm}
\usepackage{braket}
\usepackage{xcolor}
\allowdisplaybreaks
\usepackage{graphicx}

\usepackage[normalem]{ulem}

\newcommand{\figref}[1]{Fig.~\ref{#1}}

\renewcommand{\approx}{\simeq}

\usepackage[colorlinks=true]{hyperref}
\hypersetup{
    unicode=false,          
    pdftoolbar=true,        
    pdfmenubar=true,        
    pdffitwindow=false,     
    pdfstartview={FitH},    
    pdftitle={TaPtSi},    
    pdfauthor={Shashank Srivastava},      
    pdfsubject={},   
    pdfcreator={},   
    pdfproducer={}, 
    pdfkeywords={} {} {}, 
    pdfnewwindow=true,      
    colorlinks=true,       
    linkcolor=blue, 
    citecolor=blue,        
    filecolor=magenta,      
    urlcolor=blue          
}
\usepackage{orcidlink}

\begin{document}

\title{\textrm{Hourglass Dirac chains enable intrinsic topological superconductivity in nonsymmorphic silicides}}

\author{{Shashank Srivastava}\,\orcidlink{0009-0009-5065-4516}}\thanks{These authors contributed equally to this work}
\affiliation{Department of Physics, Indian Institute of Science Education and Research Bhopal, Bhopal, 462066, India}

\author{{Dibyendu Samanta}\,\orcidlink{0009-0004-3022-7633}}\thanks{These authors contributed equally to this work}
\affiliation{Department of Physics, Indian Institute of Technology, Kanpur 208016, India}

\author{Pavan Kumar Meena}
\affiliation{Department of Physics, Indian Institute of Science Education and Research Bhopal, Bhopal, 462066, India}

\author{Poulami Manna}
\affiliation{Department of Physics, Indian Institute of Science Education and Research Bhopal, Bhopal, 462066, India}

\author{Priya Mishra}
\affiliation{Department of Physics, Indian Institute of Science Education and Research Bhopal, Bhopal, 462066, India}

\author{Suhani Sharma}
\affiliation{Department of Physics, Indian Institute of Science Education and Research Bhopal, Bhopal, 462066, India}

\author{Prabin Kumar Naik}
\affiliation{Department of Physics, Indian Institute of Science Education and Research Bhopal, Bhopal, 462066, India}

\author{Rhea Stewart}
\affiliation{ISIS Facility, STFC Rutherford Appleton Laboratory, Oxfordshire, OX11 0QX, United Kingdom}

\author{Adrian D. Hillier}
\affiliation{ISIS Facility, STFC Rutherford Appleton Laboratory, Oxfordshire, OX11 0QX, United Kingdom}

\author{{Sudeep Kumar Ghosh}\,\orcidlink{0000-0002-3646-0629}}
\email{skghosh@iitk.ac.in}
\affiliation{Department of Physics, Indian Institute of Technology, Kanpur 208016, India}

\author{{Ravi Prakash Singh}\,\orcidlink{0000-0003-2548-231X}}
\email{rpsingh@iiserb.ac.in}
\affiliation{Department of Physics, Indian Institute of Science Education and Research Bhopal, Bhopal, 462066, India}

\begin{abstract}
Nonsymmorphic crystalline symmetries provide a robust route to symmetry-protected electronic topology, yet their role in stabilizing intrinsic topological superconductivity remains largely unexplored. Here, we report \ch{TaPtSi} as a new member of the superconducting nonsymmorphic silicide family, characterized via AC transport, magnetization, heat capacity, and muon spin rotation/relaxation ($\mu$SR) measurements. Zero-field $\mu$SR reveals spontaneous internal magnetic fields below $T_{\rm c}$, establishing time-reversal symmetry breaking in \ch{TaPtSi}. First-principles calculations on \ch{TaPtSi} and its isostructural nonsymmorphic superconducting analogues reveal the presence of symmetry-protected hourglass dispersions. The "necks" of these dispersions form Dirac nodal rings and chains that reside near or intersect the Fermi level. Guided by Ginzburg–Landau symmetry analysis, we identify an internally antisymmetric non-unitary triplet pairing state as the unique ground state consistent with the experimental phenomenology. Based on Bogoliubov–de Gennes calculations, we further demonstrate that this state supports Majorana surface modes, establishing its intrinsically topological nature. These results reveal a systematic route by which nonsymmorphic symmetry drives the interplay between hourglass Dirac-chain topology and unconventional triplet pairing, positioning equiatomic silicides as a unified materials platform for intrinsic topological superconductivity.
\end{abstract}

\maketitle

\section{INTRODUCTION}

Topological superconductivity represents a novel quantum phase emerging from the coexistence of superconductivity and nontrivial electronic band topology~\cite{Sato,nayak2008topo,qi2011tis}. Unlike conventional superconductors, topological superconductors are predicted to host Majorana fermions at their boundaries or defects, which are of fundamental interest for fault-tolerant quantum computation~\cite{wang2018tsm}. Nontrivial bulk band structures, enforced by crystalline symmetries, can stabilize a wide variety of topological phases~\cite{burkov2016topological,armitage2018weyl,gao2019tsm,Yadav2024}. In particular, nonsymmorphic symmetries, which combine point-group operations with fractional lattice translations, play a crucial role in protecting robust band crossings such as hourglass dispersions and Dirac nodal lines against spin-orbit coupling~\cite{wang2016nonsymmorphic,Yang2018nonsymmorphic,Biswas2021chiral,Shang2022Weyl,TSCin-non-symmorphic}. Although such symmetry-enforced topological features provide a natural setting for realizing intrinsic topological superconductivity, most experimental efforts have historically relied on proximity effects or carrier doping, which often suffer from interfacial complexity and materials-integration challenges~\cite{babio3,pbtase2}. This has motivated an intensified search for intrinsic, stoichiometric materials that simultaneously host superconductivity and symmetry-protected topological band structures. However, merely having topological bands is often insufficient; the realization of a truly topological superconducting state frequently requires unconventional pairing mechanisms that can spontaneously break time-reversal symmetry, thereby opening a gap in the bulk while leaving topologically protected states at the surface.

\begin{figure*}[!t]
\includegraphics[width=0.95\textwidth]{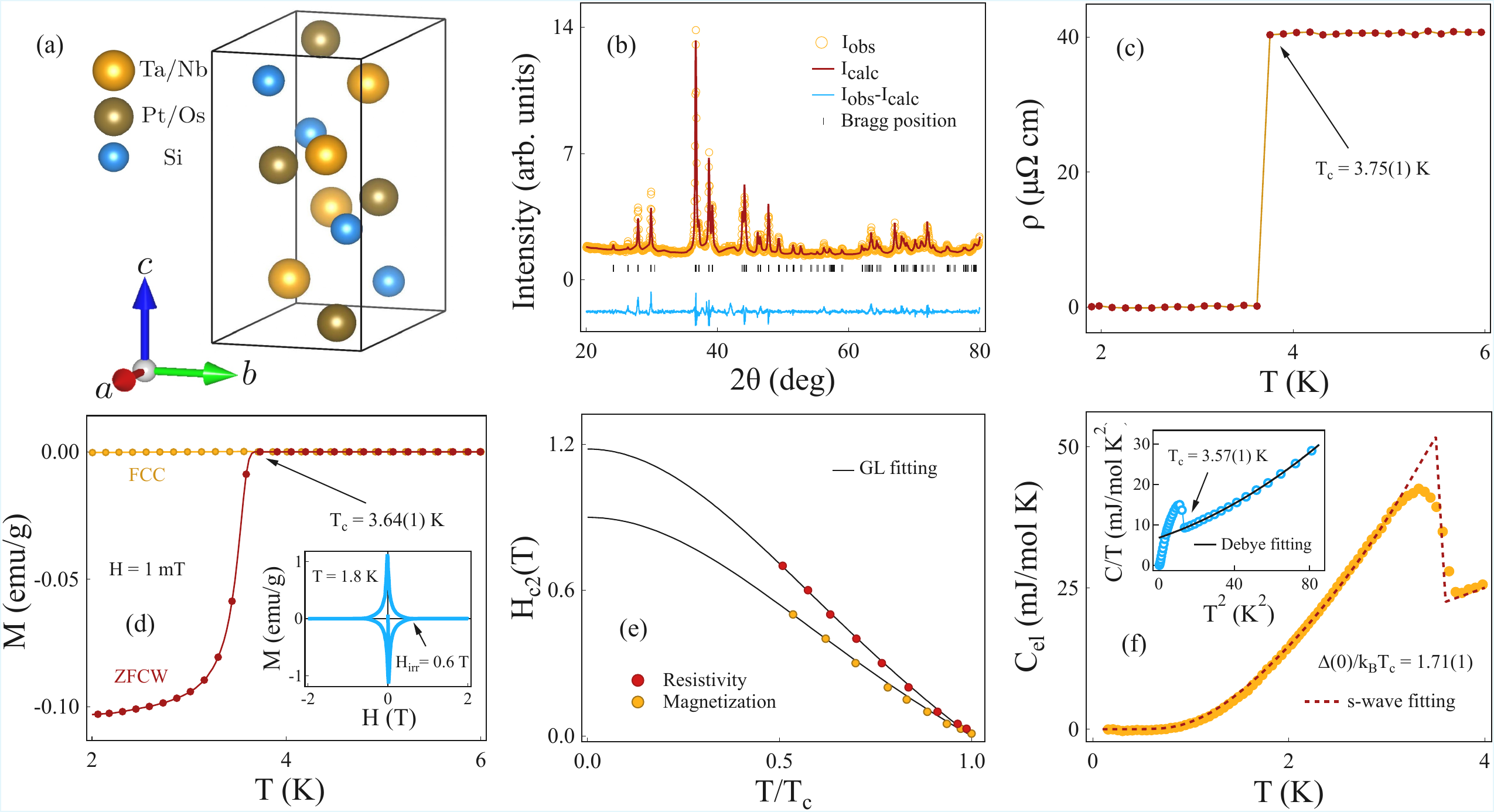}
\caption {\label{Fig1}\textbf{Structural and superconducting properties of TaPtSi:} (a) The crystal structure of $MT$Si ($M$ = Ta, Nb; $T$ = Pt, Os). (b) Rietveld refined powder XRD pattern for TaPtSi, where observed data, fitted line, their difference, and the Bragg positions are mentioned. (c) The plot of the low-temperature AC resistivity data shows a sharp zero drop in resistivity. (d) Magnetic moment versus temperature measured in ZFCW and FCC modes, showing superconductivity at 3.64(1) \si{K}. The inset displays the M-H loop at 1.8 \si{K}. (e) Variation of upper critical field with reduced temperature ($T/T_{\rm c}$), fitted with the GL equation. (f) The temperature-dependent electronic specific heat in the low-temperature region, well-fitted with the isotropic s-wave model. Inset: The data in the normal region is well described by the Debye model.}
\end{figure*}

The spontaneous break of the time-reversal symmetry (TRS) in the superconducting state is arguably one of the most promising indicators of intrinsic topological superconductivity~\cite{Sato,Ghosh2020a,kataria2026,Biswas2021chiral}. While extremely rare, superconductors that break TRS often involve complex, multi-component order parameters or nonunitary triplet pairing, which are essential ingredients for stabilizing Majorana modes~\cite{Annett1990,sigrist1991,Shang2018,Ghosh2021time,Shang2020,bhattacharyya2018unconventional,bhattacharyya2015unconventional,bhattacharyya2015broken}. In this context, superconducting topological metals, such as Dirac or Weyl semimetals, are of particular interest because the interplay between their symmetry-protected gapless excitations and TRS-breaking Cooper pairing can generate novel topological phases unavailable in standard BCS systems~\cite{armitage2018weyl,Lv2021}. Despite this potential, experimental realizations remain scarce in topological semimetals, with only a handful of examples like LaNiGa$_2$~\cite{Hillier2012,badger2022dirac,Ghosh2020b}, HfRhGe~\cite{Sajilesh2024hfrhge} and UTe$_2$~\cite{ran2019} showing clear evidence of broken TRS. Therefore, it is highly significant that spontaneous TRS breaking has recently been reported in equiatomic silicides $M$OsSi ($M$ = Ta, Nb)~\cite{Ghosh2022Dirac}. These compounds belong to the broader class of ternary transition-metal silicides crystallizing in the nonsymmorphic \ch{TiNiSi}-type structure (space group $Pnma$, No. 62)—a family known for relatively high transition temperatures and robust symmetry-protected topological band features such as Dirac nodal rings and drumhead surface states~\cite{bahadur2018mmx,zrirsi,ScRuSi,hfirsi,taptsi,Pavan_2025,sudeepZrOsSi}. However, while most members of this family preserve TRS, the behavior of $M$OsSi ($M$ = Ta, Nb) stands out as an anomaly. The microscopic origin of this TRS breaking in centrosymmetric, nonsymmorphic crystals remains unresolved~\cite{NbTaOsSifirsttrsb}, raising the critical question of whether this is an isolated phenomenon or a generic, symmetry-driven mechanism linking nonsymmorphic symmetry-enforced topology to unconventional superconductivity across the entire 111-silicide family.

In this work, we address this challenge by an extensive investigation of \ch{TaPtSi}, a new superconducting member of the nonsymmorphic 111-silicide family. Transport and thermodynamic measurements confirm bulk, fully gapped superconductivity, while zero-field $\mu$SR reveals spontaneous time-reversal symmetry breaking—establishing this as a generic feature of the $MT$Si family ($M$ = Ta, Nb; $T$ = Pt, Os). Extensive first-principles calculations combined with symmetry analysis reveal bulk hourglass dispersions in the 111-silicides, whose necks form nonsymmorphic symmetry-protected Dirac nodal chains near the Fermi level. These bulk features give rise to drumhead-like surface states, while the system also hosts well-separated Dirac topological surface states with helical spin textures that disperse across the Fermi level and remain distinct from the bulk. Guided by Ginzburg–Landau analysis, we identify an internally antisymmetric non-unitary triplet (INT) pairing state as the unique candidate explaining the coexisting full gap and broken symmetry. Calculations based on an effective minimal model demonstrate that this INT state supports Majorana surface bound states, establishing 111-silicides as a unique platform for intrinsic topological superconductivity enforced by nonsymmorphic symmetry.

\section{RESULTS AND DISCUSSION}

\subsection{Structural characterization and bulk superconducting properties of TaPtSi}
\ch{TaPtSi} polycrystalline sample synthesized by arc melting. Powder x-ray diffraction data [shown in Fig. \ref{Fig1}(b), additional details are in the Supplementary Material (SM) \cite{SM}] confirm that the sample crystallized in an orthorhombic structure with space group $Pnma$ (No. 62, point group D$_{2h}$), which is nonsymmorphic and centrosymmetric in nature. A schematic representation of the unit cell of \ch{TaPtSi} is shown in Fig. \ref{Fig1}(a). Comprehensive AC resistivity, magnetization, and specific heat measurements confirm the presence of bulk type-II superconductivity in \ch{TaPtSi}. The superconducting transition occurs at a critical temperature ($T_{\rm c}$) of 3.64(1) K, characterized by a sharp drop to zero resistivity [Fig. \ref{Fig1}(c)], a clear diamagnetic shift in the magnetization data [Fig. \ref{Fig1}(d)], and a distinct superconducting jump in the heat capacity [inset of Fig. \ref{Fig1}(f)]. The normal-state resistivity data fit well with the parallel resistor model, showing metallic behavior with a residual resistivity ratio of 4.09(1). Temperature and field-dependent magnetization and resistivity measurements were used to determine the critical fields of \ch{TaPtSi}. By fitting these data with the Ginzburg-Landau (GL) model, the zero-temperature lower critical field $H_{\rm c1}(0)$ was found to be 10.30(1) mT [SM \cite{SM}, Figure S1(b)]. The upper critical field $H_{\rm c2}(0)$ was determined to be 1.18(1) T from resistivity and 0.90(1) T from magnetization [Fig. \ref{Fig1}(e)]. These critical field values yield a GL coherence length $\xi_{GL}$ of 19.03(1) nm and a penetration depth $\lambda_{GL}$ of 198.81(1) nm. The resulting GL parameter, $\kappa_{GL}(0) = 10.40(1)$, and the thermodynamic critical field, $H_{\rm c} = 0.063(2)$ T, confirm a strong type-II superconducting nature. Additionally, the low Maki parameter ($\alpha_m = 0.14(1)$) indicates that the upper critical field is primarily governed by the orbital limiting effect rather than the Pauli limiting effect. The normal state specific heat data were analyzed using the Debye-Sommerfeld relation yielding a Debye temperature $\Theta_D = 318(1)$ K and a density of states at the Fermi energy $D_C(E_F) = 2.67(3)$ states eV$^{-1}$f.u.$^{-1}$. The weak-coupling superconductivity was confirmed from the electron-phonon coupling constant, $\lambda_{e-ph}=0.57(6)$, calculated using the inverted McMillan’s equation. The electronic-specific heat $C_{el}$ versus temperature data fit well with the isotropic s-wave model (Fig. \ref{Fig1}(f)), yielding a full superconducting gap of $\Delta(0)/{k_{B}T_{\rm c}}=1.71(1)$ [details on fittings and calculations are given in the SM \cite{SM}].

\subsection{Time-reversal symmetry breaking
from Zero field \texorpdfstring{$\mu$}{mu}SR in TaPtSi}
\begin{figure}[!t]
\includegraphics[width=\columnwidth]{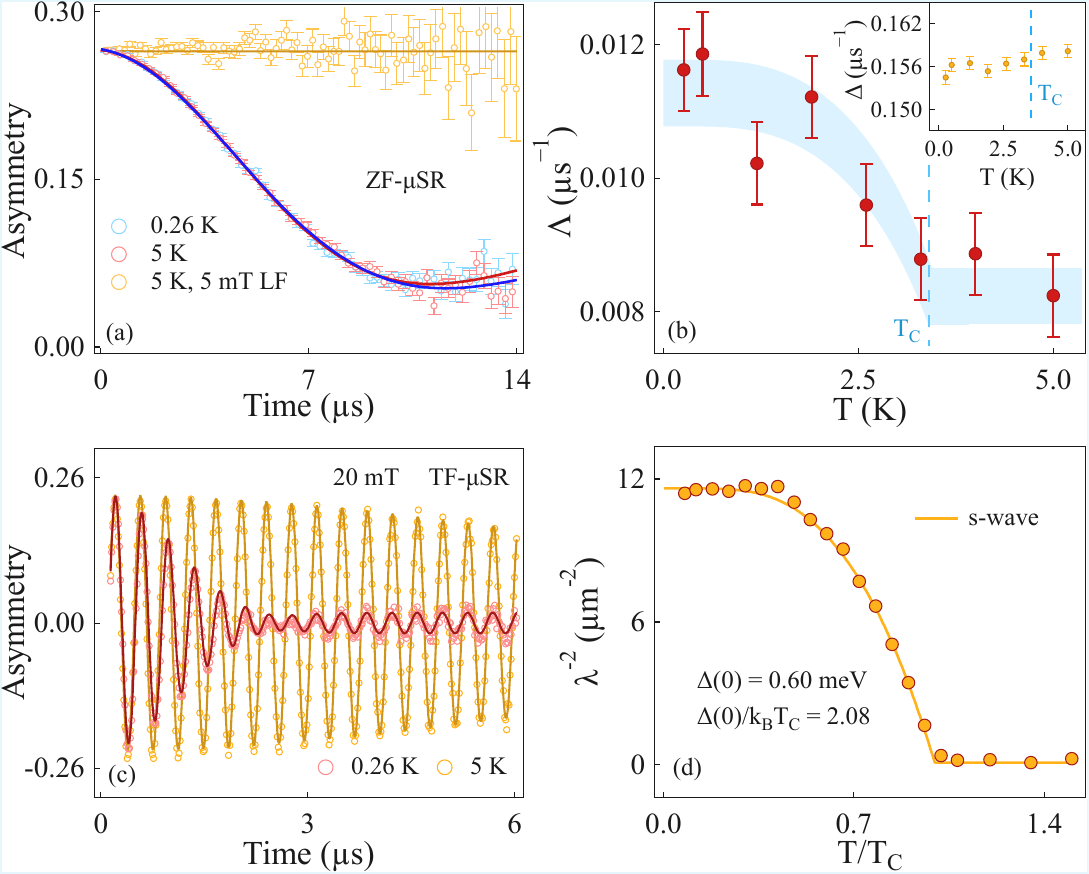}
\caption {\label{Fig3}\textbf{Time-reversal symmetry breaking with fully gapped superconductivity from $\mu$SR in \ch{TaPtSi}:} 
(a) The zero-field muon spectra recorded at 0.26 \si{K} and 5 \si{K}. Solid lines exhibit the fitting of the asymmetry spectra using Eq. \ref{Eq:kubo}. (b) The temperature variations of $\Lambda$ and $\Delta$ (Inset) along with the error bar, showing an increase in $\Lambda$ below $T_{\rm c}$, $\Delta$ being constant. (c) Temperature variation of the TF muon asymmetry spectra below $T_{\rm c}$, and above $T_{\rm c}$, under 20 \si{mT} magnetic fields. (d) The Brandt fitting yielded the fitted parameters $\lambda^{-2}$, which varied with temperature. The orange line denotes the optimal fit of the data with Eq. \ref{Eq:swavemuon}.}
\end{figure}

In the zero-field (ZF) mode of muon spin rotation/relaxation ($\mu$SR) experiments, the influence on time-reversal symmetry (TRS) caused by spontaneous magnetic fields in the superconducting state can be quite sensitively detected. The longitudinal field configuration ZF asymmetry spectra for \ch{TaPtSi} were obtained from 5 \si{K} to 0.26 \si{K} (Fig. \ref{Fig3}(a)). In the absence of static magnetic or electronic moments, the polarization of muon decay is caused by random nuclear moments. Since the plot suggests no oscillatory component, there is also no ordered magnetic structure \cite{Ghosh2020a}. The muon depolarization data can be theoretically described by the Gaussian Kubo-Toyabe function \cite{hayano1979kubo},
\begin{equation}
G_{ZF}(t) = \frac{1}{3}-\frac{2}{3}(\Delta^{2}t^{2}-1)\mathrm{exp}\left(\frac{-\Delta^{2}t^{2}}{2}\right).
\label{Eq:kubo}
\end{equation}
\noindent The relaxation function $A(t) = A_{1}G_{ZF}(t)\mathrm{exp}(-\Lambda t)+A_{BG}$ is used to fit the ZF depolarization data as shown in Fig. \ref{Fig3}(a), where $A_{1}$ and $A_{BG}$ are the asymmetry due to the sample and the time-independent background, respectively. The muon spin relaxation rates due to the nuclear and electronic moments in the material are denoted by $\Delta$ and $\Lambda$, respectively.
These relaxation rates are enhanced in the superconducting state, even in the presence of any minute spontaneous magnetic field within the detection limit of $\mu$SR \cite{Sajilesh2024hfrhge,luke1998time}. The fitting shown in the ZF spectra above and below the $T_{\rm c}$ shows a clear distinction. The fitting parameters $\Lambda$ and $\Delta$ were plotted in Fig. \ref{Fig3}(b); $\Delta$ does not show any significant change in the superconducting region, but $\Lambda$ shows a systematic enhancement below $T_{\rm c}$ that agrees with a BCS order parameter represented by the blue band \cite{re6zr,aokiprl,re6hf,singh2018re6ti}. Therefore, the superconducting ground state of \ch{TaPtSi} shows evidence of time-reversal symmetry breaking within the sensitivity of the $\mu$SR.

To ensure that the increased relaxation originates from intrinsic superconductivity rather than dilute fluctuating impurities, we conducted longitudinal field (LF)-$\mu$SR measurements. Applying 5 \si{mT} magnetic field, parallel to the initial muon-spin direction, effectively decoupled muon polarization from internal fields, resulting in an almost flat spectrum. This decoupling confirms that the relaxation observed in zero-field (ZF)-$\mu$SR is driven by static internal magnetic fields. Furthermore, the lack of oscillatory behavior in the LF data provides strong evidence against the presence of long-range magnetic order.

The magnitude of the spontaneous internal field, $|B_{int}|$, is derived from the maximum change in the nuclear relaxation rate ($\delta\Lambda$) below the superconducting transition using the relation: $|B_{int}|=\sqrt{2}\frac{\delta\Lambda}{\gamma_{\mu}}$ \cite{singh2018re6ti}. Given our experimental value of $\delta\Lambda = 0.0036(3)$, the internal field is calculated to be 0.060(2) G. This observation of time-reversal symmetry breaking (TRSB) mirrors similar findings in the isostructural compounds $M$OsSi ($M$ = Ta, Nb)~\cite{Ghosh2022Dirac}. The emergence of such a spontaneous magnetic field in \ch{TaPtSi} is particularly significant, as it suggests that high spin-orbit coupling and nontrivial topology play an unconventional role in defining the material's superconducting ground state.

\subsection{Superconducting gap structure of TaPtSi from transverse field \texorpdfstring{$\mu$}{mu}SR}
To investigate the superconducting energy gap at a microscopic level, we performed transverse-field (TF) $\mu$SR experiment. A transverse magnetic field having a constant value exceeding $H_{\rm c1}$ (0) (i.e., 20-45 \si{mT}) was applied, followed by cooling of the sample below $T_{\rm c}$. The asymmetry data are recorded above and below $T_{\rm c}$, as shown in Fig. \ref{Fig3}(c). 

The asymmetry versus time data for various temperatures and fields were fitted to obtain the variation of $\sigma$ with temperature at different fields. The parameters obtained from the fitting of the $\sigma$ versus field curves with Brandt's equation giving $\lambda^{-2}$ at different temperatures, where $\lambda$ is the penetration depth [fitting details in SM \cite{SM}]. Fig. \ref{Fig3}(d) illustrates the temperature variation of $\lambda^{-2}$, having a constant value above $T_{\rm c}$ (blue shaded region) and below $T_{\rm c}/3\approx1.2$ \si{K}. This indicates that there is no node in the superconducting gap at the Fermi surface, implying the absence of low-lying excitations \cite{la7rh3}. The fitting of $\lambda^{-2}$(T) allows us to determine the type and magnitude of the superconducting energy gap. The data align well with London's approximation for the BCS superconductors, expressed as
\begin{equation}
\frac{\lambda^{-2}(T)}{\lambda^{-2}(0)} = \frac{\Delta (T)}{\Delta (0)}\tanh{\left[\frac{\Delta (T)}{2k_BT}\right]},
\label{Eq:swavemuon}
\end{equation}
\noindent where the BCS energy gap as a function of reduced temperature is given as $\Delta({T}) = \Delta(0)\tanh[1.82(1.018(\frac{T_{\rm c}}{T}-1))^{0.51}]$ \cite{Carrington2003MgB2}. The energy gap value obtained from the fitting is 0.59(6) \si{meV}, which can be written as the BCS parameter ${\Delta(0)}/{k_{B}T_{\rm c}}=2.08(2)$. The value obtained from the TF-$\mu$SR measurement is slightly higher than the value of 1.71(1) from specific heat data, although both reveal an isotropic s-wave pairing symmetry of the energy gap. A similar discrepancy was also observed in the case of \ch{La7X3} (\ch{X} $=$ \ch{Rh}, \ch{Ir}) \cite{la7ir3,la7rh3,li2018la7ir3}. In addition, London penetration depth $\lambda_{GL}^{\mu SR}$ was found to be 293.49(6) \si{nm}. The value obtained from $\mu$SR is significantly different from the magnetization measurement. The difference, previously observed in some other compounds, although not well understood, may indicate a real difference in the penetration depth, being extracted from the Meissner state and the vortex state of the material \cite{anandadrojaRURhPAs,agsnse2,zr2ir,Pavan_2025}.

\begin{figure*}[!ht]
\includegraphics[width=0.89\textwidth]{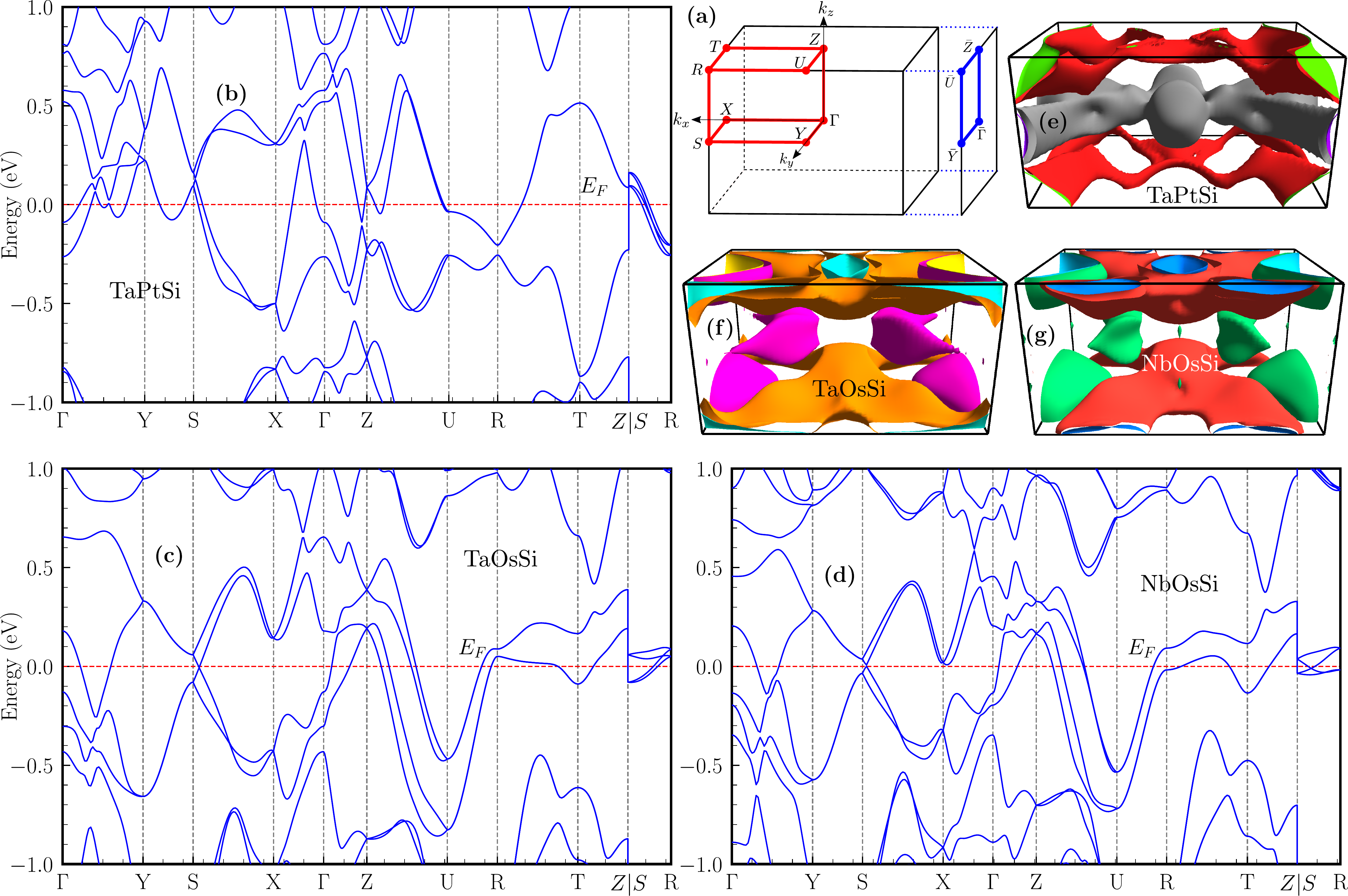}
\caption {\textbf{Electronic band structure of the silicides $MT$Si ($M$ = Ta, Nb; $T$ = Pt, Os):} (a) Three-dimensional bulk Brillouin zone (BZ) and the (100) surface BZ of the primitive orthorhombic lattice, with high-symmetry points and paths indicated by red dots and blue lines, respectively. (b,c,d) Bulk electronic band structures of TaPtSi, TaOsSi, and NbOsSi obtained with spin–orbit coupling (SOC), exhibiting hourglass-type dispersions along the $S-X$ and $S-R$ directions of the BZ. (e,f,g) Fermi surfaces of TaPtSi, TaOsSi, and NbOsSi calculated including SOC, illustrating the multiple sheets spanning the BZ.}
\label{fig:band}
\end{figure*}

\section{Nature of the superconducting ground state in 111-silicides}
Having established the bulk superconducting properties of \ch{TaPtSi} through transport, magnetization, and thermodynamic measurements, and demonstrated clear time-reversal symmetry breaking in the superconducting state from zero-field $\mu$SR while retaining a fully gapped order parameter, we now turn to a detailed theoretical investigation to elucidate the microscopic origin of this unconventional superconducting behavior. Rather than restricting the analysis to a single compound, we develop a unified theoretical framework for the broader nonsymmorphic 111 silicide family, which allows us to place the newly discovered \ch{TaPtSi} in a coherent materials context and identify family-wide organizing principles. Our theoretical analysis proceeds in a systematic manner: we first analyze the normal-state electronic structure and band topology using first-principles calculations, revealing nonsymmorphic symmetry–protected hourglass dispersions and associated Dirac nodal lines. We then employ a Ginzburg–Landau symmetry analysis appropriate for the low-symmetry $Pnma$ crystal structure to identify superconducting order parameters compatible with a fully gapped state and broken time-reversal symmetry. Finally, guided by these symmetry constraints, we construct an effective low-energy model that captures the essential hourglass band connectivity and demonstrate that the resulting superconducting state is intrinsically topological, hosting Majorana surface modes.

\subsection{Electronic band structure from first principles calculations}
Electronic structure calculations were carried out within density functional theory (DFT) using the generalized gradient approximation with the Perdew–Burke–Ernzerhof (PBE) exchange–correlation functional~\cite{Perdew_1996} to investigate the topological characteristics of the silicides $MT$Si ($M$ = Ta, Nb; $T$ = Pt, Os). These isostructural compounds crystallize in the orthorhombic space group $Pnma$ (No. $62$), a nonsymmorphic group characterized by three key symmetry operations: inversion symmetry $\mathcal{I}$ and the two glide-mirror symmetry $\mathcal{G}_x$: $\{\Pi_x|\frac{1}{2},\frac{1}{2},\frac{1}{2}\}$ and $\mathcal{G}_y$: $\{\Pi_y|\frac{1}{2},\frac{1}{2},0\}$, where $\Pi_x$ and $\Pi_y$ denote mirror reflections perpendicular to the $x$ and $y$ directions, respectively. Together with time-reversal symmetry, these crystalline symmetries generate nontrivial band crossings that underpin the topological nature of the electronic states.

\begin{figure*}
\includegraphics[width=1.9\columnwidth]{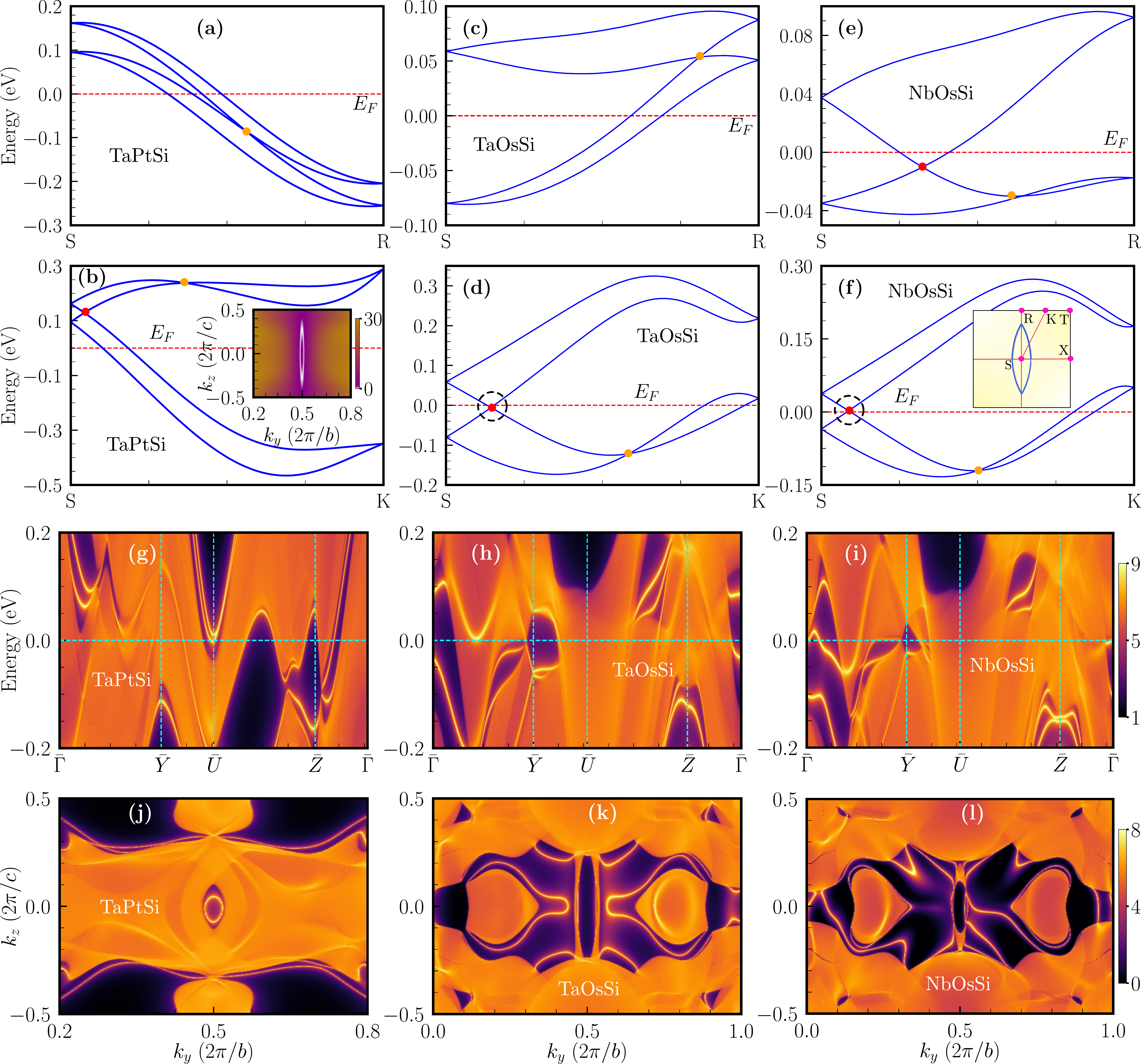}
\caption {\textbf{Hourglass dispersions, Dirac rings and surface states of $MT$Si ($M$ = Ta, Nb; $T$ = Pt, Os):} (a,c,e) Hourglass-type band dispersions of TaPtSi, TaOsSi, and NbOsSi along the high-symmetry path $S-R$. (b,d,f) Corresponding hourglass dispersions along the $S-K$ direction, where $K$ is the midpoint between $T$ and $R$ (see inset of panel (f)). Type-I and type-II Dirac crossings are highlighted by red and orange markers, respectively. The inset in (b) shows distribution of the Dirac ring (white) for TaPtSi, encircling the $S$ point. The color map indicates the magnitude of the local band gap in meV. The inset in (f) illustrates the fourfold-degenerate Dirac ring generated by the neck points (red dots) of the hybrid hourglass band structure on the $k_x=\pi$ plane. Dirac points pinned at the Fermi energy are highlighted by black dotted circles in (d) and (f). (g,h,i) Surface-state spectra along high-symmetry paths of the projected (100) surface Brillouin zone, highlighting the nontrivial drumhead surface states emerging around the $\bar{Y}$ point. (j,k,l) Constant-energy contours of the surface spectral weight for TaPtSi at –0.10 eV, TaOsSi at –0.01 eV, and NbOsSi at 0.0 eV, illustrating the momentum-space extent and morphology of the drumhead states.}
\label{fig:hourglass}
\end{figure*}

The electronic band structure calculated in the absence of spin–orbit coupling (SOC) [see SM \cite{SM}] reveals multiple dispersive bands intersecting the Fermi level, resulting in multiple electron and hole pockets and highlighting the pronounced multiband nature of these silicides. The low-energy states of TaPtSi are dominated by Ta-5d, Pt-5d, Si-3p, and Ta-5p orbitals, while in TaOsSi they primarily originate from Ta-5d, Os-5d, Si-3p, and Ta-5p contributions; for NbOsSi, the dominant states stem from Nb-4d, Os-5d, Si-3p, and Nb-4p orbitals, in decreasing order of weight, consistent with the orbital-resolved DOS presented in the Supplementary Material. Along $\Gamma$–$Y$ and $\Gamma$–$Z$ directions, red dotted circles denote the $\mathcal{G}_x$-protected band crossings that assemble into a nodal ring encircling $\Gamma$ within the $k_x = 0$ plane for TaOsSi and NbOsSi [see SM \cite{SM}]. Upon incorporation of SOC, this nodal loop becomes gapped and a band inversion emerges near $\Gamma$, giving rise to a nontrivial surface state. Although SOC does not open a global gap, a well-defined $\mathbb{Z}_2$ invariant in the $k_y-k_z$ plane ($\mathbb{Z}_2=1$) confirms the presence of robust $(100)$ surface states [see SM \cite{SM}]. In these three silicides, the SOC band structure presented in \figref{fig:band}(b–d) displays a prominent splitting near the Fermi level-most evident along the $S-X$ direction, reaching $\sim 100$ meV, arising from strong spin–orbit interactions associated with the dominant Ta-5d, Nb-4d, Os-5d and Pt-5d orbitals. The calculated Fermi surfaces in \figref{fig:band}(e–g) feature nearly parallel sheets extending throughout the Brillouin zone, providing an electronic structure conducive to enhanced interband superconducting pairing~\cite{Ghosh2022Dirac,Weng2016}.

A prominent characteristic of the SOC-influenced bands is the emergence of hourglass-shaped dispersions~\cite{gao2020r,Li2018,Pavan_2025,Pavan_2025_Zr} along the high-symmetry paths $S-R$, $S-X$, and $S-K$ (with $K$ denoting the midpoint between $T$ and $R$), as shown in \figref{fig:band}(b–d). These hourglass dispersions reflect a symmetry-enforced band connectivity in energy–momentum space, originating from the interplay between nonsymmorphic glide-mirror symmetry $\mathcal{G}_x$ and time-reversal symmetry. Because the eigenstates at different high-symmetry points transform under distinct irreducible representations, Kramers pairs must exchange partners along these directions. This enforced partner switching produces a symmetry-protected degeneracy—the characteristic “neck’’ of the hourglass—before the bands rehybridize to complete the hourglass profile. Close-up views of these features, which remain protected by $\mathcal{G}_x$, are shown in \figref{fig:hourglass}(a–f). Beyond hourglass connectivity, our analysis also reveals a Dirac ring centered at the $S$ point on the $k_x=\pi$ plane, accompanied by topologically nontrivial surface states, as illustrated in the inset of \figref{fig:hourglass}(f).

To uncover the symmetry mechanism underlying the hourglass dispersions \cite{gao2020r,Li2018}, we analyze the algebra of glide-mirror symmetry $\mathcal{G}_x$ in the $k_x=\pi$ plane. The high-symmetry line $S$–$R$, parameterized by $(\pi,\pi,k_z)$ with $-\pi<k_z<\pi$, lies entirely within this plane and remains invariant under glide operation $\mathcal{G}_x$. Along this line, the glide operation satisfies $\mathcal{G}^2_x=T_{011}\bar{\mathcal{R}}=e^{-ik_z}$, where $\bar{\mathcal{R}}$ denotes a spin rotation by $2\pi$ and $T_{011}$ represents a lattice translation along the $y$ and $z$ directions by one unit. Consequently, Bloch states are labeled by momentum-dependent glide eigenvalues $g_x=\pm e^{-ik_z/2}$. As the crystal preserves both the inversion $\mathcal{I}$ and the time-reversal symmetries $\mathcal{T}$, the Kramers partners carry identical glide eigenvalue along this path. In particular, for any Bloch state $\ket{\phi_n}$, the glide mirror $\mathcal{G}_x$ satisfies $\mathcal{G}_x(\mathcal{IT}\ket{\phi_n})=g_x(\mathcal{IT}\ket{\phi_n})$. At the time-reversal-invariant momenta $S$ and $R$, the states $\ket{\phi_n}$, $\mathcal{T}\ket{\phi_n}$, and $\mathcal{I}\ket{\phi_n}$ also share the same glide eigenvalue.

At $S=(\pi,\pi,0)$, the glide eigenvalues reduce to $g_x=\pm1$, and the degenerate quartet formed by ${\ket{\phi_n}, \mathcal{T}\ket{\phi_n}, \mathcal{I}\ket{\phi_n}, \mathcal{IT}\ket
{\phi_n}}$ separates into two  sectors distinguished by $g_x=\pm1$. Typically, the $g_x=-1$ states are energetically lower than the $g_x=+1$ states. In contrast, at $R=(\pi,\pi,\pi)$ the eigenvalues become $g_x=\pm i$, and each Kramers pair is formed by states carrying opposite glide eigenvalues. As a result, the fourfold-degenerate manifold at $R$ consists of two states with $g_x=+i$ and two with $g_x=-i$. The mismatch between the glide eigenvalue structure at $S$ and $R$ forces a rearrangement of band partners along the path $S-R$, producing the characteristic hourglass dispersion. Notably, this band connectivity is dictated entirely by the nonsymmorphic glide symmetry, rendering the crossing symmetry-enforced and robust. An analogous eigenvalue evolution occurs along the $S$–$X$ and $S$–$K$ directions, where glide symmetry likewise protects hourglass dispersions. In these latter cases, dispersions span a broader energy window, whereas the feature along $S$–$R$ remains confined to a narrow energy range.

The hourglass dispersions of the silicides $MT$Si ($M$ = Ta, Nb; $T$ = Pt, Os), shown in \figref{fig:hourglass}(a–f), host glide-enforced hourglass band connectivity featuring both type-I and type-II Dirac crossings, indicated by red and orange markers, respectively. The type-I Dirac nodes occur at the “neck’’ of the hourglass dispersion and emerge systematically along the $S$–$K$ direction, where $K$ denotes an arbitrary point on the $T$–$R$ line; analogous crossings also appear along the $R$–$X$ direction. These symmetry-enforced neck points continuously connect to generate a Dirac nodal ring encircling the $S$ point within the $k_x=\pi$ plane, as illustrated schematically in the inset of \figref{fig:hourglass}(f). Notably, in the case of NbOsSi the Dirac nodal ring is weakly warped, with substantial portions lying at or very close to the Fermi level, rendering it highly accessible to experimental probes. This nonsymmorphic symmetry-driven Dirac ring is corroborated by our first-principles DFT calculations, explicitly shown for TaPtSi in the inset of \figref{fig:hourglass}(b), with the corresponding results for TaOsSi and NbOsSi presented in the Supplementary Material.

Finally, we analyze the surface electronic structure by projecting the bulk bands onto the (100) surface Brillouin zone, as shown in \figref{fig:hourglass}(g-i). The surface spectra reveal several topologically nontrivial states that cross the Fermi level. These arise from the bulk Dirac nodal loops, whose projections onto the surface Brillouin zone give rise to characteristic drumhead-like features~\cite{Li2018}. The absence of inversion symmetry on the surface leads to a noticeable splitting of these drumhead states~\cite{Li2018}. The corresponding surface bands, tied to the underlying Dirac rings, appear near $\sim -0.11$ eV in TaPtSi, $\sim -0.01$ eV in TaOsSi and $\sim 0.0$ eV in NbOsSi, as illustrated in \figref{fig:hourglass}(j-l).

\begin{figure}[!t]
\includegraphics[width=0.98\columnwidth]{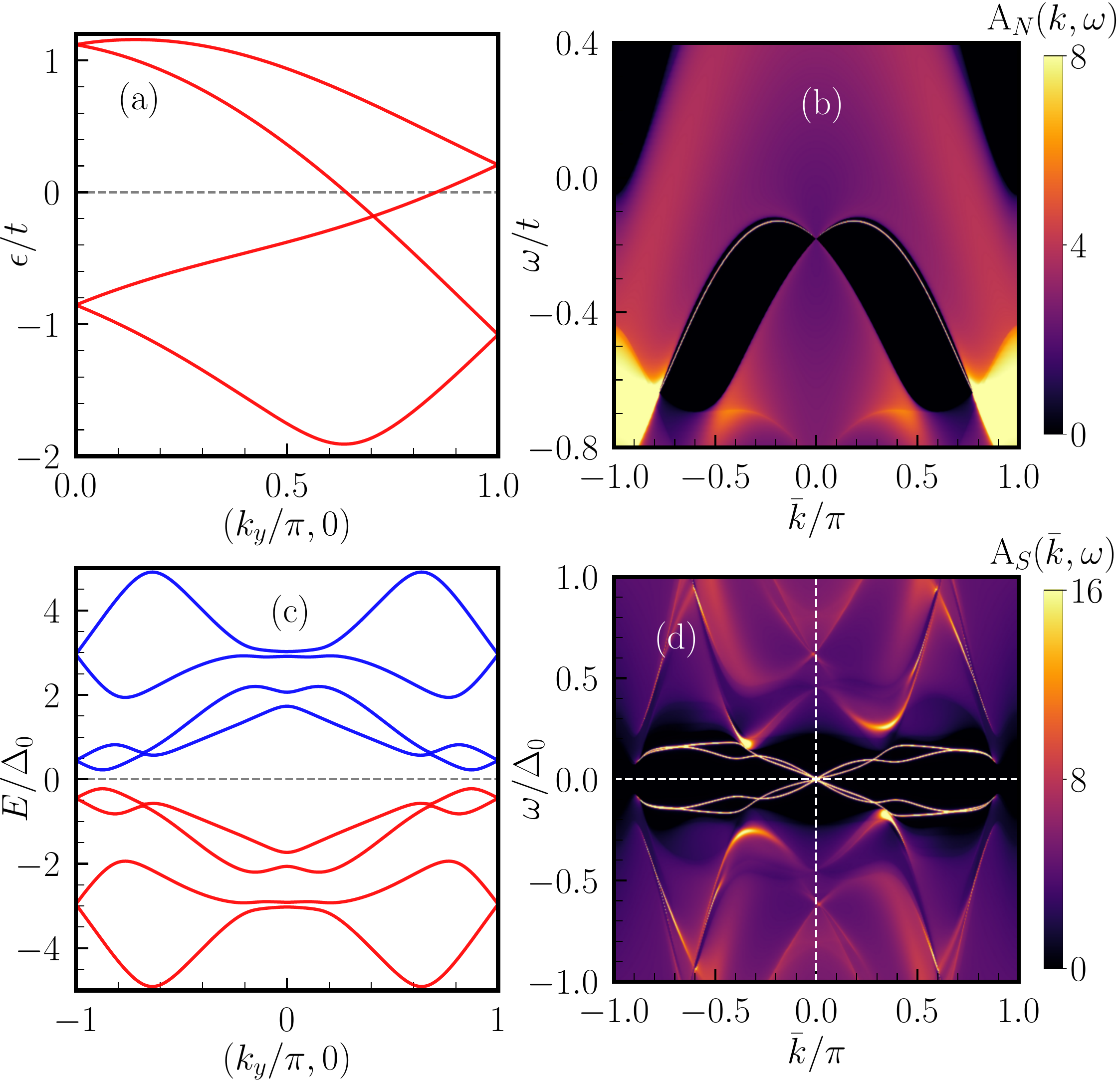}
\caption {\label{fig:tsc} \textbf{Topological superconductivity enabled by hourglass topology:} (a) Normal-state band dispersion exhibiting the hourglass feature along the $k_y$ direction for a minimal model capturing Dirac hourglass connectivity. (b) Surface spectral function of the Dirac hourglass semimetal represented by the minimal model, with the color scale representing the surface spectral weight, A$_N(\bar{k},\omega)$. (c) Bogoliubov–de Gennes quasiparticle spectrum of the internally antisymmetric nonunitary triplet (INT) superconducting state in the hourglass model, showing a fully gapped superconducting state. (d) Surface spectral function of the hourglass semimetal in the INT superconducting phase, A$_S(\bar{k},\omega)$, revealing linearly dispersing Majorana surface modes. Here, $\omega$ denotes the surface energy, and $\bar{k}$ represents the momentum along the $\hat{z}$ direction of the surface Brillouin zone. The model parameters used are $t_a=0.5t$, $t_c=0.5t$, $\lambda=0.8t$, $t_\perp=0.7t$, $\lambda_\perp=0.4t$, $m_1=0.6t$, $m_2=0.8t$, $\lambda_{\perp}^\prime=0.3t$, and $\mu=3.0t$. For the INT pairing, the triplet vector components are chosen as $\eta_x=\tfrac{1}{\sqrt{2}}$, $\eta_y=\tfrac{e^{i\pi/4}}{\sqrt{2}}$, and $\eta_z=0$.}
\end{figure}

\subsection{Topological superconducting ground state}
To uncover the nature of the superconducting ground state of the 111-silicides $MT$Si ($M$ = Ta, Nb; $T$ = Pt, Os), we begin with a symmetry-based analysis within the Ginzburg–Landau framework~\cite{Ghosh2020a,Annett1990,sigrist1991}. These materials crystallize in the low-symmetry orthorhombic point group $D_{2h}$, which admits only one-dimensional irreducible representations. Consequently, in the strong spin–orbit coupling (SOC) limit, symmetry forbids any superconducting instability that spontaneously breaks time-reversal symmetry (TRS). In the weak SOC regime, where the normal state possesses $D_{2h} \otimes \mathrm{SO}(3)$ symmetry, TRS-breaking superconducting states are in principle allowed; however, all such states necessarily host point nodes~\cite{Hillier2012,Weng2016}, which is incompatible with the experimentally observed fully gapped superconductivity in these silicides. Our first-principles calculations indicate substantial SOC effects, further ruling out these nodal TRS-breaking scenarios. Although nonsymmorphic symmetries can enforce nodal superconductivity along high-symmetry directions at the Brillouin-zone boundary, they do not generate the multicomponent order parameters required for TRS breaking~\cite{sumita2018}. Therefore, all symmetry-allowed single-band superconducting instabilities in $MT$Si are inconsistent with the experimental observations of the fully gapped superconducting state with spontaneous TRS breaking.

The multiband character of silicides $MT$Si revealed by our first-principles calculations naturally motivates us to consider an internally antisymmetric nonunitary triplet (INT) superconducting state—an unconventional pairing mechanism previously proposed for LaNiGa$_2$~\cite{Hillier2012,Weng2016,badger2022dirac,Ghosh2020b}, a compound that shares the same point group $D_{2h}$. In this pairing state, electrons on the same atomic site form Cooper pairs across different orbitals, with fermionic antisymmetry enforced in the orbital space. The pairing potential takes the form $\hat{\Delta}=(i\tau_y)\otimes(\mathbf{d}\cdot\mathbf{s})(i s_y)$, where the triplet vector is $\mathbf{d}=\Delta_0\boldsymbol{\eta}$ with $|\boldsymbol{\eta}|^2=1$ and $\Delta_0$ the uniform pairing amplitude. The state becomes nonunitary when $\mathbf{q}=i(\boldsymbol{\eta}\times\boldsymbol{\eta}^*)\neq0$, indicating the spin-selective pairing that naturally breaks TRS. Such an order parameter can arise from local interband attraction mediated by Hund’s coupling. Moreover, the presence of nearly parallel and quasi-degenerate Fermi-surface sheets shown in \figref{fig:band}(e,f,g) provides favorable conditions for stabilizing the INT superconducting state in these 111-silicides.

We next investigate the interplay between the nonsymmorphic symmetry–protected bulk hourglass dispersion in the $k_y$–$k_z$ plane [\figref{fig:hourglass}(a–f)], whose Dirac neck points lie in close proximity to the Fermi level, and the INT superconducting pairing. To isolate the essential low-energy physics associated with the Dirac hourglass band connectivity, we construct an effective minimal model Hamiltonian defined in the $k_y$–$k_z$ plane. We begin by defining a one-dimensional hourglass ladder along the $k_y$ direction, where kinetic terms generate a Dirac-like backbone and specific asymmetry terms induce the partner switching of Kramers pairs that forms the characteristic hourglass “neck”. Crucially, this 1D sector includes a nonsymmorphic spin-orbit coupling (SOC) that intertwines spin, sublattice, and chain degrees of freedom to protect the crossing. We then extend this to a two-dimensional hourglass semimetal by coupling these ladders via hopping along $k_z$. This inter-ladder coupling introduces dispersion along the transverse direction and additional SOC components that shape the spin texture of the bands. By adjusting mass and offset terms, we can precisely tune the position of the resulting Dirac node relative to the Fermi level, while subleading SOC interactions lift residual degeneracies to stabilize the low-energy spectrum for the emergence of superconductivity. The resulting dispersion of the minimal model explicitly reproduces the characteristic hourglass profile along the $k_y$ direction, as shown in \figref{fig:tsc}(a), while the remaining degrees of freedom encode the nonsymmorphic symmetry constraints (see SM \cite{SM} for details). This minimal framework enables a controlled analysis of how the hourglass feature imprints nontrivial topology onto the INT superconducting state. 

To implement INT pairing, the required orbital antisymmetry can originate from the node, sublattice, or chain subspaces. Evaluating these options within the Bogoliubov-de Gennes formalism, we find that only the sublattice-based pairing $\hat{\Delta}=\kappa_0\otimes(\mathbf{d}\cdot\mathbf{s})(is_y)\otimes(i\sigma_y)\otimes\tau_0$ yields a fully gapped quasiparticle spectrum, as shown in \figref{fig:tsc}(c). The alternative choices invariably result in nodal spectra that contradict the experimental evidence. Thus, the sublattice-based INT channel is a unique solution capable of stabilizing the nodeless superconducting state observed in these silicides. We further investigate the topological character of this superconducting phase by evaluating the surface spectral properties of the state by computing the retarded surface Green’s function $G(\bar{k},\omega)$ using the transfer-matrix method [see SM \cite{SM} for details]. The corresponding surface spectral function is defined as $A(\bar{k},\omega)=-\frac{1}{\pi}Im\left[ \operatorname{Tr}G(\bar{k},\omega) \right]$. For reference, the spectral function of the normal hourglass state ($A_N(\bar{k},\omega)$) is shown in \figref{fig:tsc}(b), while the spectral function of the superconducting state ($A_S(\bar{k},\omega)$) is presented in \figref{fig:tsc}(d). The appearance of zero-energy surface Andreev bound states (SABS) shown in \figref{fig:tsc}(d) demonstrates that the INT superconducting state in these silicides is topologically nontrivial. Notably, a linearly dispersing Majorana mode emerges at the $\Gamma$ point. Moreover, the SABS exhibit a characteristic twisting dispersion that connects the Majorana branch at $\Gamma$ to a secondary crossing away from $\Gamma$, revealing the interplay between nonsymmorphic symmetry and nonunitary triplet pairing~\cite{Hsieh_2012,Hao_2011}.

Finally, we highlight that the symmetry-protected topological surface states in 111-silicides, shown in \figref{fig:hourglass}(g,h,i), are energetically isolated from the bulk continuum and traverse the Fermi level. This separation allows the bulk superconducting condensate to induce a distinct superconducting gap on the surface via the proximity effect, establishing these materials as a promising platform for surface topological superconductivity as well. The resulting two-gap structure—a primary bulk gap coexisting with a smaller induced surface gap—can be resolved experimentally using Andreev reflection spectroscopy, as recently demonstrated in YRuB$_2$~\cite{mehta2024yrub2}, or via high-resolution angle-resolved photoemission spectroscopy (ARPES)~\cite{Yang2023}.

\section{Summary and Conclusions}
We have synthesized and characterized the new nonsymmorphic superconductor \ch{TaPtSi} ($T_{\rm c} \approx 3.64$~K) and demonstrated the coexistence of a fully isotropic superconducting gap and spontaneous time-reversal symmetry (TRS) breaking. While thermodynamic and transport measurements consistently indicate weak-coupling, fully gapped superconductivity, zero-field $\mu$SR measurements reveal spontaneous internal magnetic fields below $T_{\rm c}$. First-principles calculations, supplemented by symmetry analysis, demonstrate that nonsymmorphic crystal symmetries stabilize hourglass band dispersions whose necks form spin--orbit-protected Dirac nodal rings and chains near the Fermi level. These nodal structures generate both drumhead-like surface states and robust Dirac topological surface states with helical spin textures, that disperse across the Fermi level and remain well separated from the bulk. We show that this distinctive electronic structure provides the microscopic environment required for stabilizing unconventional superconductivity. Within a Ginzburg--Landau framework tailored to the $Pnma$ space group, we identify an INT pairing state as the only symmetry-allowed order parameter compatible with the observed phenomenology. This inter-orbital triplet state intrinsically breaks time-reversal symmetry while maintaining a fully open quasiparticle gap, thereby reconciling thermodynamic isotropy with the spontaneous internal fields detected by $\mu$SR. Furthermore, our Bogoliubov--de Gennes calculations based on an effective model establish that the resulting superconducting ground state is intrinsically topological and supports zero-energy Majorana bound states at the surface.

As nonsymmorphic symmetry-protected hourglass metals that are also bulk superconductors, 111-silicides provide a uniquely versatile platform for exploring topological quantum phenomena. On the one hand, they realize intrinsic topological superconductivity in which the bulk condensate itself generates Majorana boundary modes; on the other, their normal state hosts robust helical surface states that offer a natural template for proximity-induced superconductivity~\cite{Li2018}. The coexistence of strong spin--orbit coupling, helical spin textures, and unconventional pairing can enable electrically controllable spin-polarized supercurrents relevant for superconducting spintronics~\cite{Ma2017experimental,Eschrig2015}. In addition, the presence of bulk Dirac nodal loops facilitates unconventional Josephson junctions with anomalous current--phase relations~\cite{Parhizgar2020} and provides a promising route toward non-Hermitian superconducting platforms through engineered dissipation~\cite{Cayao2024,Shen2024}. Collectively, these features establish the 111-silicide family as a novel testbed for investigating the interplay of nonsymmorphic symmetry, electronic topology, and unconventional superconductivity in future quantum devices.

\section{Methods}

\noindent\textit{Experimental details.} High-purity pieces of Ta (4N), Pt (4N), and Si (5N) were used in the stoichiometry for the synthesis. They were placed on the water-cooled copper hearth under an argon atmosphere. The metals were melted to form an alloy and the resulting button was repeatedly remelted to increase homogeneity. The melted button was annealed for 7 days at $1100$\si{\celsius}. Powder X-ray diffraction (XRD) was performed at room temperature using a PANalytical X’Pert diffractometer to confirm the phase purity. A Quantum Design superconducting quantum interference device (SQUID) magnetometer (MPMS-3, 7T) was used to measure the magnetization in vibrating sample magnetometer (VSM) mode. Electric transport measurement was performed using a physical property measurement system (PPMS, 9T) in a four-probe configuration. The specific heat was measured using the two-tau relaxation mode with the PPMS equipped with a dilution refrigerator (DR).\\

\noindent\textit{$\mu$SR measurements.} Measurements of muon spin rotation and relaxation have been performed in the MuSR facility at the ISIS neutron and muon source, RAL, STFC, UKRI, United Kingdom \cite{isisdoi}. The 3\si{g} powder of the \ch{TaPtSi} sample was prepared in the silver holder using diluted GE varnish and mounted on a sorption cryostat. An active field compensator system using current-carrying coils ensured a zero field, thus eliminating the effect of external stray fields (see \cite{hillier2022muon} for details).\\

\noindent\textit{Electronic band structure and topology.}
Electronic structure calculations were carried out within density functional theory (DFT) using the \textsc{QUANTUM ESPRESSO} package~\cite{Giannozzi_2009,Giannozzi_2017,Giannozzi_2020}. The generalized gradient approximation (GGA) with the Perdew–Burke–Ernzerhof (PBE) exchange–correlation functional~\cite{Perdew_1996} was used, together with projector augmented-wave (PAW) pseudopotentials. A plane-wave kinetic energy cutoff of 80~Ry was used and the Brillouin zone was sampled using a mesh of points $\Gamma$-centered $8 \times 10 \times 8$ $k$. The experimental lattice parameters and atomic positions obtained from the Rietveld refinement were adopted without further structural relaxation. Maximally localized Wannier functions were constructed using the \textsc{WANNIER90} package~\cite{Pizzi2020}. The resulting Wannier-based tight-binding Hamiltonian was subsequently used to compute the Fermi surfaces and surface spectral functions of a semi-infinite SrPd$_2$As$_2$ slab via the iterative Green’s function method implemented in \textsc{WannierTools}~\cite{Wu2018}.\\

\noindent\textit{Toy model for Dirac hourglass dispersion.}
We develop an effective low-energy model for the hourglass Dirac semimetal identified in our first-principles calculations. Earlier studies of nonsymmorphic hourglass fermions have focused primarily on hourglass nodal-line and hourglass Weyl semimetals~\cite{Wang_2017}. Here, we construct a minimal Hamiltonian tailored to an hourglass Dirac semimetal, capturing the essential symmetry-enforced band connectivity. The model is formulated in two steps, beginning with a one-dimensional hourglass ladder along the $k_y$ direction, which serves as the fundamental building block of the full model:
\begin{align}
H_{\mathrm{HG}}^{(1D)}(k_y) &= h^{(1D)}(k_y) \kappa_z;\nonumber\\
\!\!\!\!\!\!\!\!\!\! h^{(1D)}(k_y) &= t~(1 + \cos k_y) \sigma_x + t~\sin k_y \sigma_y + t_c \tau_x \nonumber \\
&+ t_a ~\bigg[(1 - \cos k_y) \sigma_x \tau_z - \sin k_y \sigma_y \tau_z \bigg] \nonumber \\
&+ \lambda s_y \sigma_z \tau_y\,,
\end{align}
Physically, the first two terms ($\propto t$) represent the kinetic energy within the sublattice sector, generating the Dirac-like dispersion that forms the backbone of the hourglass structure. The term proportional to $t_c$ hybridizes the two chain degrees of freedom, introducing the necessary multiband character. Crucially, the asymmetry terms ($\propto t_a$) couple the sublattice and chain sectors to control the partner switching between Kramers pairs—this mechanism fixes the location and shape of the characteristic hourglass “neck”. The final term ($\propto \lambda$) represents a nonsymmorphic spin-orbit coupling (SOC) that intertwines spin, sublattice, and chain degrees of freedom; this term is symmetry-allowed by the glide mirror and is essential for protecting the hourglass crossing against gap opening.

To obtain the full two-dimensional hourglass semimetal, we couple these ladders via symmetry-allowed interladder hopping along the $k_z$ direction:
\begin{align}
\!\!\!\!\!\!\!\!\!\!\!\! H_{\mathrm{HG}}^{(2D)}(k_y,k_z) &= [h^{(1D)}(k_y) + h_\perp(k_z)]\kappa_z; \nonumber\\
h_\perp(k_z) &= t_\perp (\cos k_z \tau_x - \sin k_z \tau_y) \nonumber \\
&+ \lambda_\perp (\cos k_z s_y \sigma_z \tau_y + \sin k_z s_y \sigma_z \tau_x) \nonumber \\
&+ \frac{1}{2} (m_1 + m_2) + \frac{1}{2} (m_1 - m_2) \sigma_z \tau_z \nonumber\\
&+ \lambda_{\perp}^\prime \sin k_z s_x \sigma_z \tau_z
\end{align}
Here, the interladder hopping term ($\propto t_\perp$) generates dispersion along $k_z$, promoting the system from a 1D ladder to a 2D semimetal. The corresponding SOC term ($\propto \lambda_\perp$) is critical for shaping the spin texture of the resulting Dirac bands. The terms involving $m_1$ and $m_2$ function as energy offset and mass parameters, allowing us to tune the relative position of the Dirac crossings and place the Dirac nodal ring  with respect to the Fermi level. Finally, the subleading SOC term ($\propto \lambda_{\perp}^\prime$) mixes additional spin components; while it does not destroy the hourglass dispersion, it lifts residual degeneracies, which is vital for stabilizing a fully gapped superconducting state when pairing is introduced.

\section*{Acknowledgments} 
Sh.~S. acknowledges University Grants Commission (UGC), Government of India, for the Senior Research Fellowship (SRF). R.~P.~S. acknowledges the ANRF Government of India for the Core Research Grant No. CRG/2023/000817 and ISIS, STFC, UK, for providing beamtime for the $\mu$SR experiments. S.~K.~G. acknowledges financial support from Anusandhan National Research Foundation (ANRF) erstwhile Science and Engineering Research Board (SERB), Government of India via the Startup Research Grant: SRG/2023/000934 and from IIT Kanpur via the Initiation Grant (IITK/PHY/2022116). S.~K.~G. and D.~S. utilized the \textit{Andromeda} server at IIT Kanpur for numerical calculations.

\bibliography{Library}

\clearpage

\onecolumngrid

\begin{center}
    {\Large \textbf{\textrm{Supporting Information for\\ "Hourglass Dirac chains enable intrinsic topological superconductivity in nonsymmorphic silicides"}}}
\end{center}

\vspace{10pt}


\setcounter{figure}{0} 

\renewcommand{\thefigure}{S\arabic{figure}}  

\setcounter{table}{0} 

\renewcommand{\thetable}{S\arabic{table}}  


In the supporting information, we present details on the structural characterization, resistivity, magnetization, specific heat, and muon spin relaxation/rotation experimental data for \ch{TaPtSi}, as well as additional details on the theoretical calculations of the nonsymmorphic 111-silicide family $MT$Si ($M$ = Ta, Nb; $T$ = Os, Pt).

\section*{Characterization and Superconducting properties}
\subsection*{Structural characterization}
The FullProf Suite software \cite{fullprof} was used to refine the powder XRD pattern using the Rietveld approach. The crystal structure was generated using the VESTA software \cite{momma2011vesta}. The lattice constants of $a$ = 6.4061(7) $\text{\AA}$, $b$ = 3.8053(3) $\text{\AA}$, $c$ = 7.3796(5) $\text{\AA}$, and $V_{cell}$ = 179.89(3) $\text{\AA}^{3}$ were extracted from the Rietveld refinement and found to be in agreement with the previously reported values \cite{taptsi}.

\subsection*{Electrical Resistivity}
Temperature-dependent resistivity $\rho(T)$, as shown in Fig. \ref{Fig:S1}(a), is measured under zero field from 1.9 \si{K} to 300 \si{K}. The sharp zero drop in resistivity (Fig. \ref{Fig1}(c)) confirms superconductivity with an onset transition temperature $T_{\rm c}$ of 3.75(1) \si{K} in the TaPtSi superconductor. The temperature width of the zero-drop is less than 0.13 \si{K}, and the residual resistivity ratio ($RRR$), defined as $\rho(300K)/\rho(5K)$, is 4.09(1). The values of the transition width and $RRR$ for the polycrystalline sample indicate a high quality of our superconducting sample \cite{taptsi}. The resistivity versus temperature data above the superconducting region follows saturating behavior, which can be modeled using Wiesmann's parallel resistor formalism \cite{parallel}. The model states that temperature-dependent total resistivity $\rho(T)$ consists of two terms of resistivity, including limiting resistivity $\rho_{max}$ at high temperature and temperature-variable ideal resistivity $\rho_{i}$, formulated by Eq. \ref{Eq:Parallel}.
\begin{equation}
\frac{1}{\rho(T)}= \frac{1}{\rho_{i}(T)}+ \frac{1}{\rho_{max}},
\label{Eq:Parallel}
\end{equation}

\noindent where, $\rho_{i}(T)$ = $\rho_{i,L}(T)$ + $\rho_{i,0}$. Wilson defined the terms $\rho_{i,0}$ as the temperature-independent residual resistivity corresponding to the impurity scattering, and the $\rho_{i,L}(T)$ represents the temperature-dependent resistivity contribution caused by the electron-phonon scattering \cite{parallel2}, defined as

\begin{equation}
{\rho_{i,L}(T)}= A_0 \left(\frac{T}{\Theta_{R}}\right)^{n_0} \int_{0}^{{\Theta_{R}}/T} \frac{x^{n_0}}{(1-e^{x})(e^{-x}-1)} dx.
\label{Eqn:Parallel}
\end{equation}

\noindent $A_0$ and $\Theta_{R}$ symbolize a material-dependent variable and the Debye temperature, respectively, while $n_0$ generally has a magnitude of 2, 3, or 5 depending on the nature of interaction \cite{parallel3}. The data for the samples mentioned were best fitted with $n_0=3$ (Fig. \ref{Fig:S1}(a)), giving the values of $\rho_{max}$, $\rho_{i,0}$ and $\Theta_{R}$ as 352.0(5) \si{\micro\ohm\cdot\cm}, 45.9(4) \si{\micro\ohm\cdot\cm}, and 184(2) \si{K}, respectively.

\subsection*{Magnetization}
To measure the magnetization data, two approaches have been applied: zero field-cooled warming (ZFCW) followed by field-cooled cooling (FCC) at an applied field of 1 \si{mT}, to confirm bulk superconductivity in the TaPtSi compound. A strong diamagnetic transition in ZFCW below 3.64(1) \si{K} is shown in Fig. \ref{Fig1}(d), having $\sim100\%$ superconducting volume fraction. Splitting in the ZFCW and FCC curves is due to strong magnetic-field pinning, suggesting type-II superconductivity in the sample. Magnetization versus magnetic field loop between $\pm2$ \si{T} at 1.8 \si{K} further affirms the type-II superconductivity with $H_{irr}\approx0.6$ \si{T} (inset of Fig. \ref{Fig1}(d)).

\begin{figure*}[t]
\includegraphics[width=0.72\columnwidth]{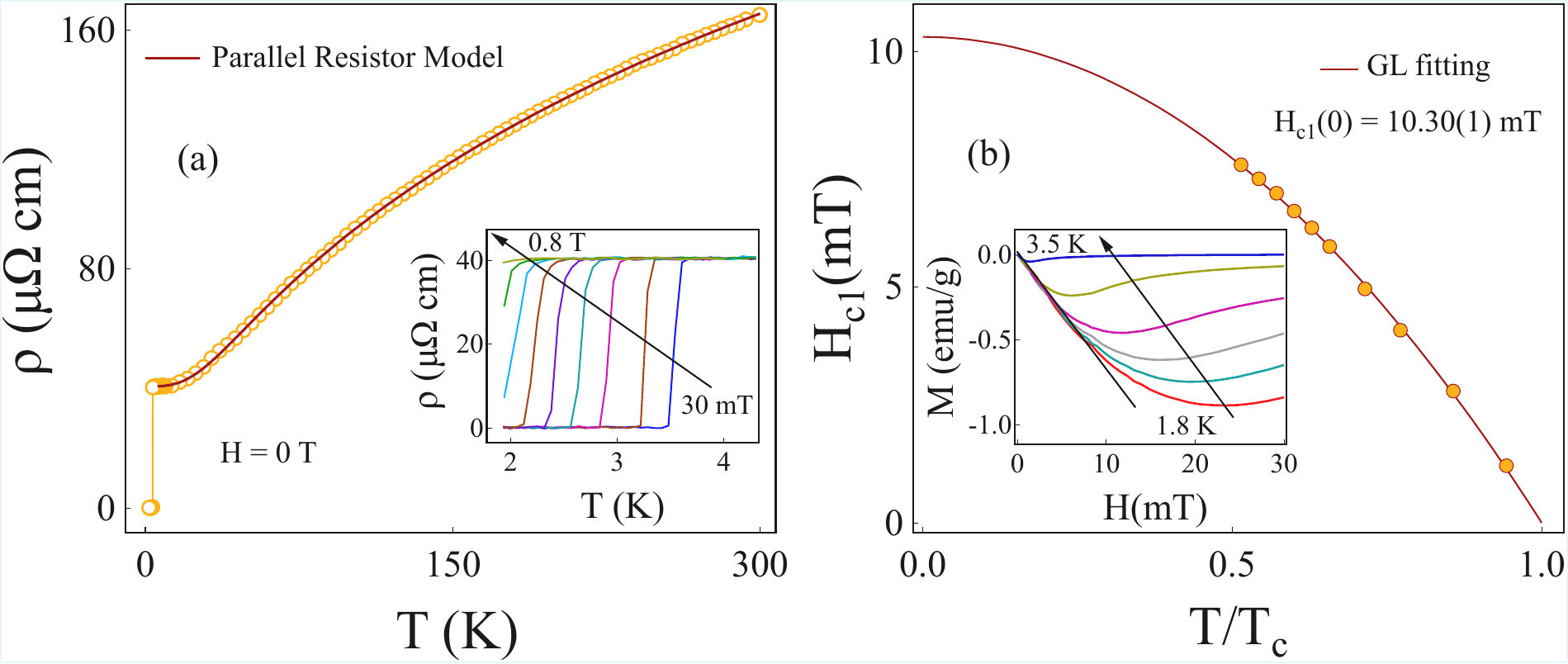}
\caption {\label{Fig:S1}(a) AC resistivity versus temperature $\rho(T)$ under zero magnetic field fitted with a parallel resistor model. Inset: The plot for the temperature-dependent resistivity at different applied fields. (b) Variation of lower critical field with reduced temperature ($T/T_{\rm c}$), fitted with the Ginzburg-Landau (GL) equation. Insets show the low magnetic-field-dependent magnetic moment at various temperatures.}
\end{figure*}

At various temperatures below $T_{\rm c}$, the magnetization was measured in the low magnetic field range (see the inset of Fig. \ref{Fig:S1}(b)). At each temperature, the lower critical fields $H_{\rm c1}$ are determined by extracting the value of the magnetic field where the curves diverge from the Meissner line. The Ginzburg-Landau (GL) equation (Eq. \ref{eqn:HC1,HC2}) was used to fit the variation of $H_{\rm c1}$ with the reduced temperature, $t$ ($=T/T_{\rm c}$), as shown in Fig. \ref{Fig:S1}(b).

\begin{equation}
H_{\rm c1}(T)=H_{\rm c1}(0)(1-t^{2}),\quad H_{\rm c2}(T) = H_{\rm c2}(0)\left[\frac{1-t^{2}}{1+t^{2}}\right].
\label{eqn:HC1,HC2}
\end{equation}

\noindent For \ch{TaPtSi}, $H_{\rm c1}(0) = 10.30(1)$ \si{mT} is obtained by intersecting the extrapolated fitting curve with the y-axis. The temperature variation of both the resistivity and magnetization data can be used to get the values of the upper critical field, $H_{\rm c2}(T)$. The resistivity and magnetization curves provide the $H_{\rm c2}$ values at different temperatures by indicating a decrease in $T_{\rm c}$ as the magnetic field is raised (Fig. \ref{Fig:S1}(a) inset). Similar to $H_{\rm c1}$, the GL equation described by Eq. \ref{eqn:HC1,HC2} (see Fig. \ref{Fig1}(e)) also provided a good fit to the values of the upper critical field, $H_{\rm c2}$ versus reduced temperature. The value of $H_{\rm c2}(0)$ derived from resistivity and magnetization measurements is 1.18(1) \si{T} and 0.90(1) \si{T}, respectively.

There are two mechanisms that suppress superconductivity in the presence of a magnetic field: the spin paramagnetic effect and the orbital limiting effect. The disruption of Cooper pairs is caused by the Zeeman splitting in the spin paramagnetic effect, exceeding the Pauli limiting field $H_{\rm c2}^{P}$(0), which equals $1.86\times T_{\rm c}$ \cite{Chandrasekhar1962pauli, Clogston1962pauli}. When the kinetic energy of the superelectrons surpasses the superconducting gap energy, the Cooper pair is believed to be split due to the orbital limiting effect. The orbital limiting field $H^{orbital}_{\rm c2}$(0) was formulated by Eq. \ref{eqn5:WHH} for type-II superconductors by the Werthamer-Helfand-Hohenberg (WHH) theory \cite{WHH1966orbital, Helfand1966orbital}.

\begin{equation}
H^{orbital}_{\rm c2}(0) = -\alpha T_{\rm c} \left.{\frac{dH_{\rm c2}(T)}{dT}}\right|_{T=T_{\rm c}}, 
\label{eqn5:WHH}
\end{equation}

\noindent The calculated values of $H^{orbital}_{\rm c2}$(0) and $H_{\rm c2}^{P}$(0) are 0.67(5) \si{T} and 6.77(1) \si{T}, respectively, for the purity factor, $\alpha=0.693$. The $H_{\rm c2}$(0) for \ch{TaPtSi} is considerably lower than the $H_{\rm c2}^{P}$(0), signifying a substantial orbital contribution. The Maki parameter, $\alpha_{m}=\sqrt{2}\frac{H_{\rm c2}^{orbital}(0)}{H_{\rm c2}^{P}(0)}$, quantifies the orbital effect in the Cooper pair breaking. The value of the Maki parameter is 0.14(1), ensuring a significant orbital contribution.

\begin{table}[t]
\caption{Normal and superconducting parameters for \ch{TaPtSi} compound, extracted from resistivity, magnetization, specific heat, and $\mu$SR measurements.}
\label{tbl:parameters}
\setlength{\tabcolsep}{35pt}
\renewcommand{\arraystretch}{1.3} 
\begin{center}
\resizebox{0.7\columnwidth}{!}{
\begin{tabular}{lcc}\hline \hline
Parameters                                  & unit                  & \ch{TaPtSi}  \\
\hline
$T_{\rm c}^{mag}$                               & \si{K}                & 3.64(1)       \\
$H_{\rm c1}(0)$                                 & \si{mT}               & 10.30(1)       \\ 
$H_{\rm c2}^{res}$(0)                           & \si{T}                & 1.18(1)       \\
$H_{\rm c2}^{P}$(0)                           & \si{T}                & 6.77(1)       \\
$\xi_{GL}$                                  & \si{nm}               & 19.03(1)       \\
$\lambda_{GL}^{mag}$                        & \si{nm}               & 198.81(1)      \\
$\lambda_{GL}^{\mu SR}$                        & \si{nm}               & 293.49(6)      \\
$\kappa_{GL}$(0)                               & -                     & 10.40(1)       \\
$\rho_{300K}/\rho_{5K}$                    & -                     & 4.09(1)       \\
$\gamma_{n}$                                & \si{mJmol^{-1}K^{-2}} & 6.30(4)      \\
$\Theta_{D}$                           & \si{K}                & 318(1)        \\
$\frac{\Delta(0)}{k_{B}T_{\rm c}}$ (sp. heat)   &  -                    & 1.71(1)       \\
$\frac{\Delta(0)}{k_{B}T_{\rm c}}$ ($\mu$SR)    &  -                    & 2.08(2)       \\
$\lambda_{e-ph}$                            & -                     & 0.57(6)       \\
$D_{C}(E_{F})$                              & states/(eV f.u.)      & 2.67(3)       \\
$m^{*}/m_{e}$                               & -                     & 1.57(7)        \\
$n_s$                                         & 10$^{26}$ \si{m^{-3}}       & 5.17(7)       \\
$T_F$                                       & -                & 1718(17)     \\
\hline \hline
\end{tabular}
}
\par\medskip\footnotesize
\end{center}
\end{table}

There are two characteristic superconducting length parameters, the coherence length $\xi_{GL}$ and the penetration depth $\lambda_{GL}$, which can be calculated using the critical fields. On solving the equations $H_{\rm c2}(0) = {\frac{\phi_{0}}{2\pi \xi_{GL}^2(0)}}$ ($\phi_{0}$ is the magnetic flux quantum), and $H_{\rm c1}(0) = \frac{\phi_{0}}{4\pi\lambda_{GL}^2(0)}\left[ln \frac{\lambda_{GL}(0)}{\xi_{GL}(0)} + 0.12\right]$, the values of $\xi_{GL}$ and $\lambda_{GL}$ were found to be 19.03(1) \si{nm} and 198.81(1) \si{nm}, respectively \cite{Cava2007coherence, tinkham2004introduction}.

Superconductors can be categorized into type-I and type-II classes based on the ratio of $\lambda_{GL}$ to $\xi_{GL}$, referred to as the Ginzburg-Landau (GL) parameter $\kappa_{GL}$(0). The value of $\kappa_{GL}$(0) for \ch{TaPtSi} is 10.40(1), significantly above $1/\sqrt{2}$, thereby identifying it as a strong type-II superconductor. 
Additionally, $H_{\rm c1}$(0), $H_{\rm c2}$(0), and $\kappa_{GL}$(0) can be used to evaluate the thermodynamic critical field parameter $H_{\rm c}$ \cite{tinkham2004introduction}. The formula $H_{\rm c}^2 ln\kappa_{GL}(0) = {H_{\rm c1}(0) H_{\rm c2}(0)}$ yields $H_{\rm c}$ = 0.063(2) \si{T}. Table \ref{tbl:parameters} lists all of the superconducting parameters estimated for the \ch{TaPtSi} compound.

\begin{figure*}[t]
\includegraphics[width=0.75\columnwidth]{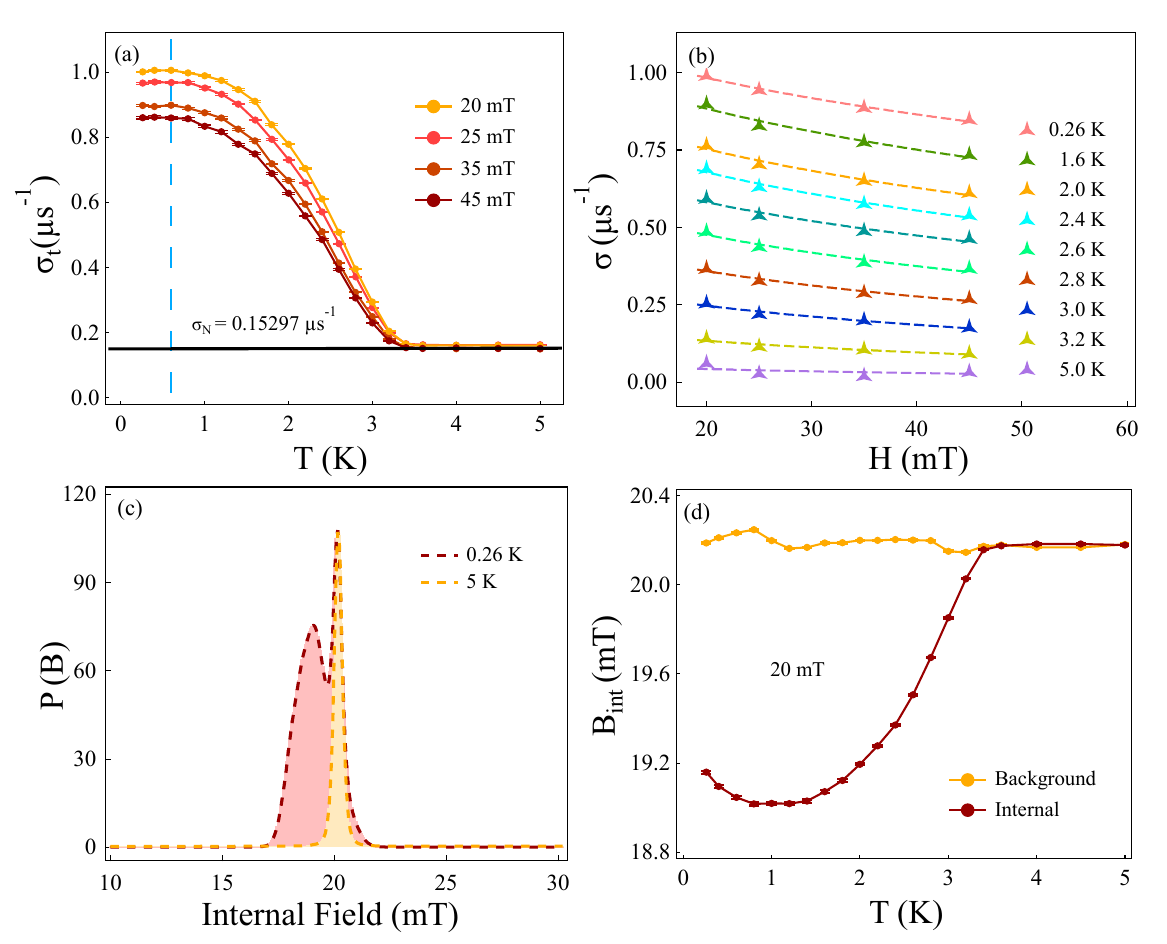}
\caption {\label{Fig:S2}(a) Temperature-dependent muon depolarization rate $\sigma$, at applied magnetic fields ranging from 20 to 45 \si{mT}, obtained from the TF-$\mu$SR data. The dashed cyan line indicates the isothermal variation of $\sigma$ with the applied field. (b) Eq. \ref{Eq:Brandt} was used to fit the variation of spin depolarization rate with magnetic field for varying temperatures. The colored dashed line represents the fitting at various temperatures. (c) Maximum entropy (MaxEnt) field distribution of the TF spectra below and above $T_{\rm c}$. (d) The temperature dependence of the internal magnetic field.}
\end{figure*}

\subsection*{Specific Heat}
The inset of Fig. \ref{Fig1}(f) shows the temperature dependence of specific heat $C$ using $C/T$ versus $T^{2}$ plot for the \ch{TaPtSi} compound under zero magnetic field. The jump in specific heat at 3.57(1) \si{K} transition temperature ensures bulk superconductivity. In a normal state, data were fitted using the Debye-Sommerfeld relation given as $C=\gamma_{n}T+\beta_{3}T^{3}+\beta_{5}T^{5}$, where the first, second, and third terms represent the contribution from electrons, the lattice, and the anharmonicity of the lattice. The obtained values corresponding to the coefficient are given as $\gamma_{n}$ = 6.30(4) \si{mJmol^{-1}K^{-2}}, phononic constants $\beta_{3}$, and  $\beta_{5}$ are 0.18(1) \si{mJmol^{-1}K^{-4}} and 0.001(1) \si{mJmol^{-1}K^{-6}}, respectively. Further Sommerfeld coefficient and phononic constant were used to calculate the density of state at the Fermi energy $D_{C}(E_{F})$ and Debye temperature $\Theta_{D}$ from the given formulas as $\gamma_{n} = \left(\frac{\pi^{2} k_{B}^{2}}{3}\right)D_{C}(E_{F})$, where $k_{B}=1.38 \times 10^{-23}$ \si{JK^{-1}} and $\Theta_{D}=\left(\frac{12\pi^{4} R N}{5 \beta_{3}}\right)^{\frac{1}{3}}$. Here, $N$ is the number of atoms per formula unit and $R=8.314$ \si{Jmol^{-1}K^{-1}}. The value of $\Theta_{D}$ obtained here can be used to calculate the electron-phonon coupling constant $\lambda_{e-ph}$ using the inverted McMillan’s equation \cite{mcmillan1968transition}:

\begin{equation}
\lambda_{e-ph} = \frac{\mu^{*}\mathrm{ln}(\Theta_{D}/1.45T_{\rm c})+1.04}{(1-0.62\mu^{*})\mathrm{ln}(\Theta_{D}/1.45T_{\rm c})-1.04}.
\label{eqn2:Lambda}
\end{equation}

\noindent Taking the repulsive-screened Coulomb pseudopotential parameter, $\mu^{*}$ as 0.13 for intermetallic systems, $\lambda_{e-ph}$ is calculated to be 0.57(6). This indicates that a weakly coupled electron-phonon interaction governs superconductivity in this system.

There are models relating the temperature-dependent superconducting energy gap $\Delta(t)$ to the electronic specific heat $C_{el}$, which are used to determine the nature of superconducting gap symmetry. The Eq. \ref{Eq:swave} relates the BCS superconducting energy gap $\Delta(t) = tanh[1.82\{1.018(\frac{1}{t}-1)\}^{0.51}]$ to the entropy $S_{el}$, which can be further linked to the electronic-specific heat by $C_{el}=t\frac{dS_{el}}{dt}$, where $t=\frac{T}{T_{\rm c}}$. So, to get the $C_{el}$, the phononic and the anharmonic contributions need to be subtracted from the total specific heat $C$ using the Debye relation. The normalized value of the jump in the electronic specific heat is $\Delta C_{el}/{\gamma_{n}T_{\rm c}}=1.29$, which is less than the BCS value (i.e., 1.43) for weakly coupled superconductors, indicating weak coupling in \ch{TaPtSi}. The BCS approximation gives rise to the relation for the normalized entropy as,

\begin{equation}
\begin{split}
\frac{S_{el}}{\gamma_{n} T_{\rm c}}= -\frac{6}{\pi^{2}} \left(\frac{\Delta(0)}{k_{B} T_{\rm c}}\right) &\int_{0}^{\infty}[ f_{y}ln(f_{y})+(1-f_{y})ln(1-f_{y})]dy,
\end{split}
\label{Eq:swave}
\end{equation}

\noindent where $y=\xi/\Delta(0)$ and $f_{y}(\xi)=[1+e^{\frac{E(\xi)}{k_{B}T}}]^{-1}$ is the Fermi function. The Fermi energy is related to the normal electron energy $E(\xi)$ by $E(\xi)=\sqrt{\xi^{2}+\Delta^{2}(t)}$. The obtained value of $C_{el}$ versus temperature was found to fit well with the isotropic s-wave model, as shown in Fig. \ref{Fig1}(f) \cite{padamsee1973quasiparticle}. From the fitting, the superconducting energy gap value was found to be $\Delta(0)/{k_{B}T_{\rm c}}=1.71(1)$, which is comparable to the ideal BCS superconducting gap of 1.76. So, the values reported in this section provide compelling evidence of weakly coupled isotropic s-wave superconductivity in \ch{TaPtSi}. Therefore, the thermodynamic measurements indicate a conventional, weakly coupled superconducting state, which is odd in the unconventional TRS-breaking superconductor.

\subsection*{\texorpdfstring{Transverse field $\mu$}{mu}SR Analysis}
The TF-$\mu$SR asymmetry spectra are accurately fitted by the function comprising the sum of sinusoidal oscillations, suppressed by a Gaussian relaxation term (Fig. \ref{Fig3}(c)) \cite{maisuradze2009comparison,weber1993flux}:

\begin{equation}
G(t) = \sum_{i=1}^N A_{i}\exp\left(-\frac{1}{2}\sigma_i^2t^2\right)\cos(\gamma_\mu B_it+\phi).
\label{Eq:TF}
\end{equation}

\noindent Here $\gamma_{\mu}/2\pi=135.5$ \si{MHz/T} is the muon gyromagnetic ratio. The Gaussian relaxation rate, the initial phase, the initial asymmetry, and the $i^{th}$ term of the Gaussian field distribution are denoted by $\sigma_i$, $\phi$, $A_i$, and $B_i$, respectively. The data fits best with two Gaussian components ($N=2$) and $\sigma_{2}=0$ for the background contribution of non-depolarizing muons stopped at the sample holder. The parameter $A_2$ denotes the background asymmetry, and $B_2$ is the magnetic field asymmetry.

\begin{figure}[t]
\includegraphics[width=0.5\columnwidth]{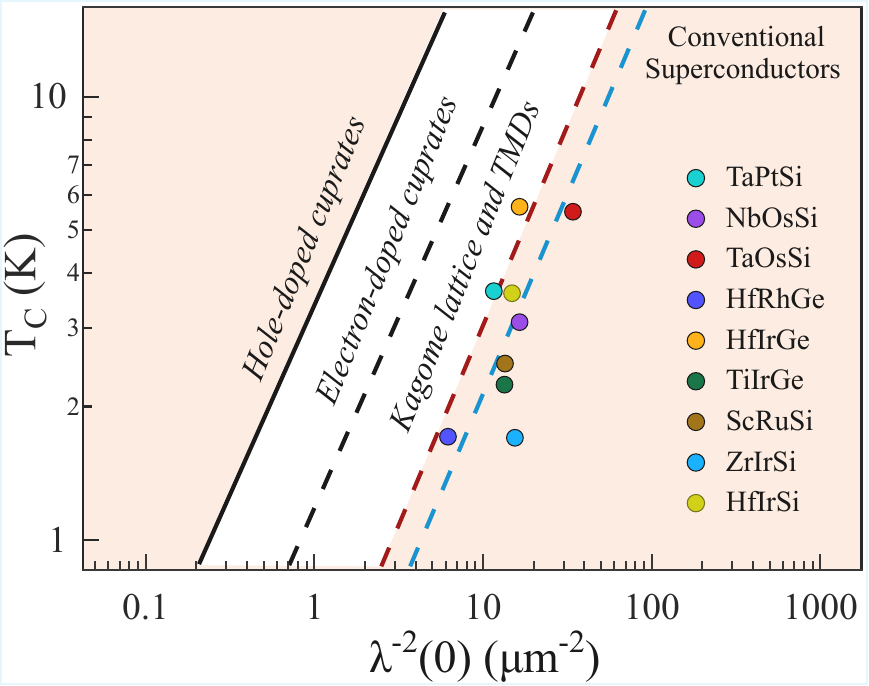}
\caption {\label{Fig6} Uemura plot based on the ratio of superconducting critical temperature ($T_{\rm c}$) to superfluid density ($\lambda^{-2}(0)$). Solid black, dashed black, and dashed brown lines indicate the ratio obtained for hole-doped cuprates, electron-doped cuprates, and transition metal dichalcogenides (TMDs) and Kagome compounds, respectively. The dashed blue line represents a guide to the nearly linear region followed by the TiNiSi-structured superconductors, close to the TMDs and Kagome superconductors \cite{Ghosh2022Dirac,Sajilesh2024hfrhge,Pavan_2025,ScRuSi,zrirsi,hfirsi}.}
\end{figure}

The variation of total Gaussian relaxation rate $\sigma_{t}$ is plotted against the temperature for the measured field between 20 and 45 \si{mT} with error bars (Fig. \ref{Fig:S2}(a)). The FLL component of the Gaussian relaxation rate $\sigma$ can be deduced by subtracting the nuclear dipolar induced temperature-independent background contribution $\sigma_N$ (shown in Fig. \ref{Fig:S2}(a)) from the $\sigma_{t}$ using $\sigma = \sqrt{\sigma_{t}^{2} - {\sigma_N}^{2}}$. The magnetic field variation of the obtained $\sigma$ at various temperatures is shown in Fig. \ref{Fig:S2}(b), inferred from the isothermal lines (indicated by the cyan line in Fig. \ref{Fig:S2}(a)). These isothermal values can be fitted with Brandt's equation \cite{Brandt} to relate the penetration depth $\lambda$ and the magnetic field for a type-II superconductor, with $\kappa_{GL}>5$. Having reduced magnetic field $h = H/H_{\rm c2}(T)$, the equation is given as,

\begin{equation}
\sigma(\mu s^{-1}) = 4.854 \times 10^{4}(1-h)[1+1.21(1-\sqrt{h})^{3}]\lambda^{-2}.
\label{Eq:Brandt}
\end{equation}

The obtained values of temperature-dependent $\lambda^{-2}$ from the above fitting were fitted with the s-wave model, as discussed in the main text (Fig. \ref{Fig3}(d)). Thus, the TF-$\mu$SR also suggests a weakly coupled s-wave superconductivity, which is peculiar in the superconductor with spontaneous magnetic field.

Fig. \ref{Fig:S2}(c) represents the magnetic field distribution in the mixed state and the normal state, respectively, acquired from the maximum entropy (MaxEnt) analysis of the Fourier-transformed TF-$\mu$SR spectra \cite{maxent}. The narrow peak in both distributions at the applied field corresponds to the muons that stop at the silver sample holder and normal-state regions of the superconductors. The wider peak at the lower field can be noticed in Fig. \ref{Fig:S2}(c) for the superconducting state, which is absent for the normal state distribution. The peak in the field less than the applied field arises for type-II superconductors in the mixed state, owing to the formation of the quantized flux line lattice.

\begin{figure}[t]
\includegraphics[width=\columnwidth]{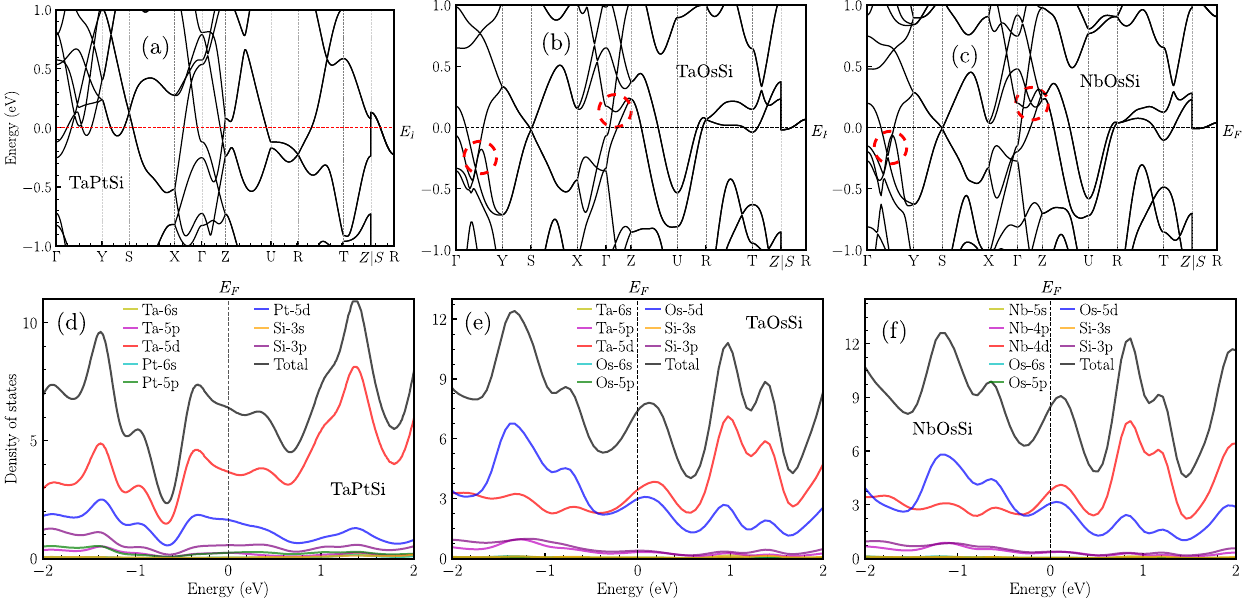}
\caption {\label{fig:bands_pdos} (a-c) Electronic band structure without including the spin-orbit coupling for the TaPtSi, TaOsSi, and NbOsSi. (d-f) Orbital resolved electronic density of states without spin-orbit coupling for TaPtSi, TaOsSi, and NbOsSi.}
\end{figure}

\noindent The temperature variation of the internal magnetic field was extracted from the fitting of TF-$\mu$SR asymmetry data at an applied magnetic field of 20 \si{mT} as presented in Fig. \ref{Fig:S2}(d). Due to Meissner field expulsion, the internal magnetic field in the superconducting state is less than the applied field. However, above $T_{\rm c}$, it corresponds to the applied field and is consistent with the background magnetic contribution, which remains constant throughout the recorded temperature range.

The $\lambda_{e-ph}$ (from specific heat data) and $\lambda_{GL}^{\mu SR}$ can be used to calculate the values of effective mass, $m^*=(1+\lambda_{e-ph})m_e$ and superconducting electron density, $n_s(0)=m^*/\mu_0e^2(\lambda_{GL}^{\mu SR})^2$, which were found to be 1.57(7)$m_e$ and $5.17(7) \times10^{26}$ \si{m^{-3}}, respectively. Furthermore, Eq. \ref{eqn:tf} can be implemented to determine the Fermi temperature ($T_F$) \cite{hillier1997classification}.

\begin{equation}
k_{B}T_{F} = \frac{\hbar^{2}}{2m^{*}}{(3\pi^{2}n_s)^{2/3}}.
\label{eqn:tf}
\end{equation}

\noindent Table \ref{tbl:parameters} lists the values that were obtained pertaining to each of these electronic parameters. Uemura et al. formulated a way to distinguish unconventional superconductors from conventional ones \cite{Uemura}. If the ratio of $T_{\rm c}$ and $T_F$ is between 0.1 and 0.01, the superconductor is in the unconventional region. The ratio of $T_{\rm c}$ and $T_F$ for \ch{TaPtSi} was found to be 0.00212, which lies outside the unconventional regime, which is puzzling for a TRS-breaking superconductor.

The linear trend in the ratio between superconducting critical temperature, $T_{\rm c}$, and the superfluid density, $\lambda^{-2}(0)$, was first observed for cuprates by Uemura et al. \cite{Uemura}. Such a linear relationship may exist even for systems with $T_{\rm c}$/$T_F$ values 20 times lower than those of hole-doped cuprates ($\sim$0.05) \cite{guguchia2017signatures,guguchia2023unconventional,TMDsUemura}. The ratio between $T_{\rm c}$ and $\lambda^{-2}(0)$ for the reported TiNiSi-structured compounds along with \ch{TaPtSi} lies in the vicinity of the dashed blue line (Fig. \ref{Fig6}). The nearly linear trend indicates that the compounds have a shared pairing mechanism and an unconventional electronic origin of superconductivity.

\section*{Band structure and topology}

The band structure computed without spin–orbit coupling (SOC) exhibits multiple dispersive bands crossing the Fermi level, resulting in several electron and hole pockets and highlighting the strongly multiband nature of these silicides as shown in Figure~\ref{fig:bands_pdos}(a-c). In TaPtSi, the low-energy electronic states are primarily derived from Ta-5d, Pt-5d, Si-3p, and Ta-5p orbitals. For TaOsSi, the dominant contributions originate from Ta-5d, Os-5d, Si-3p, and Ta-5p states, whereas in NbOsSi the low-energy spectrum is governed mainly by Nb-4d, Os-5d, Si-3p, and Nb-4p orbitals, in decreasing order of weight. These orbital characters are consistent with the orbital-resolved density of states presented in Figure~\ref{fig:bands_pdos}(d-f). Along the $\Gamma-Y$ and $\Gamma-Z$ directions, $\mathcal{G}_x$-protected band crossings appear and connect to form a nodal ring encircling the $\Gamma$ point in the $k_x=0$ plane for TaOsSi and NbOsSi. Upon including SOC, this nodal loop becomes gapped and a band inversion develops near $\Gamma$, signaling a transition to a nontrivial topological phase.

\begin{figure}[t]
\includegraphics[width=0.75\columnwidth]{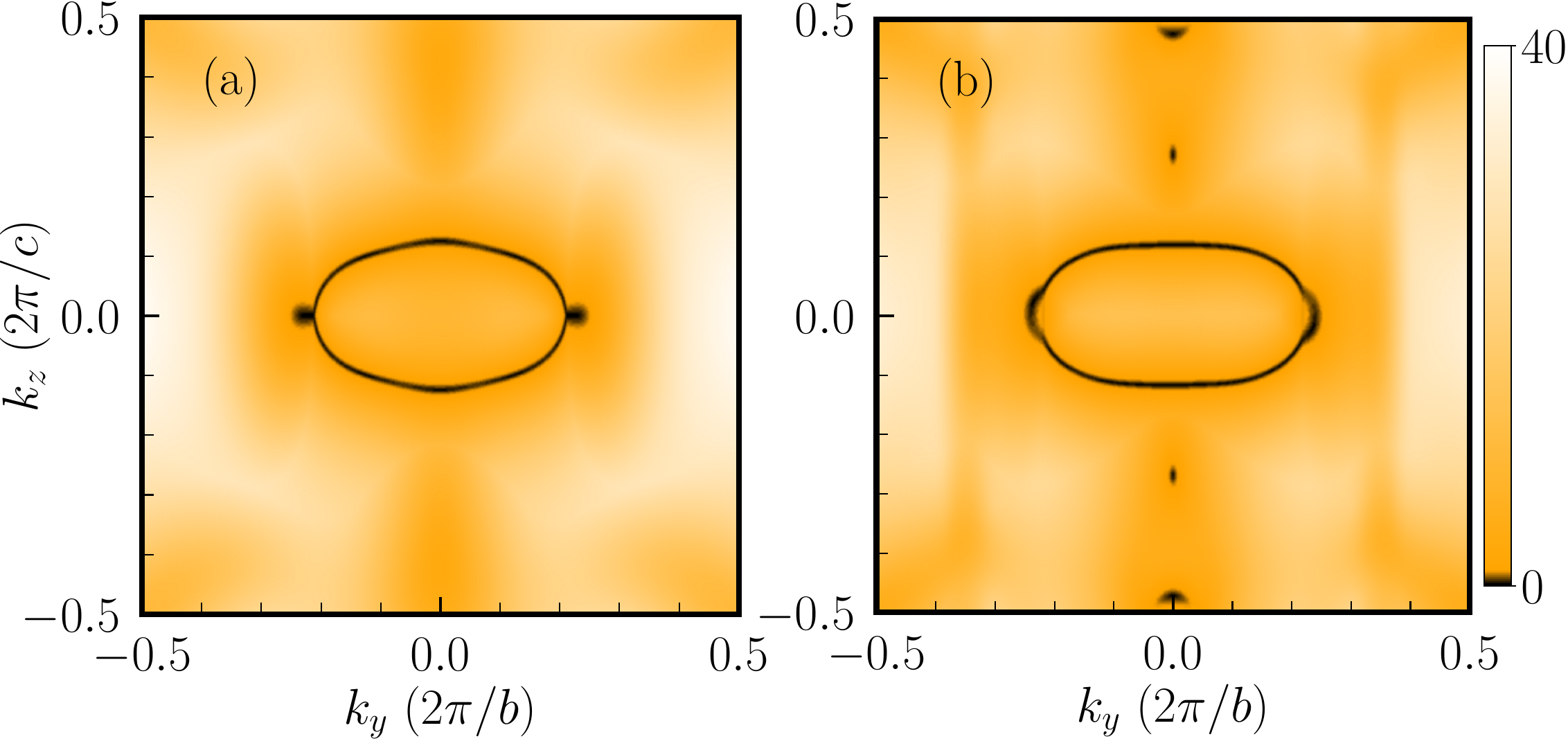}
\caption {\label{fig:nodal_loop} The nodal ring centered at the $\Gamma$ point in the $k_x = 0$ plane for (a) TaOsSi and (b) NbOsSi, calculated without SOC.}
\end{figure}

In the absence of SOC, our calculations thus reveal a symmetry-protected Dirac nodal ring centered at $\Gamma$ within the $k_x=0$ plane. The resulting nodal loops for TaOsSi and NbOsSi are shown in Figure~\ref{fig:nodal_loop}(a) and \ref{fig:nodal_loop}(b), respectively. This topological feature is enforced by the nonsymmorphic glide-mirror symmetry $\mathcal{G}_x:(x,y,z)\rightarrow(-x+\tfrac{1}{2},y+\tfrac{1}{2},z+\tfrac{1}{2})$, which guarantees the stability of the band crossings forming the nodal ring.

\begin{figure}[t]
\includegraphics[width=0.75\columnwidth]{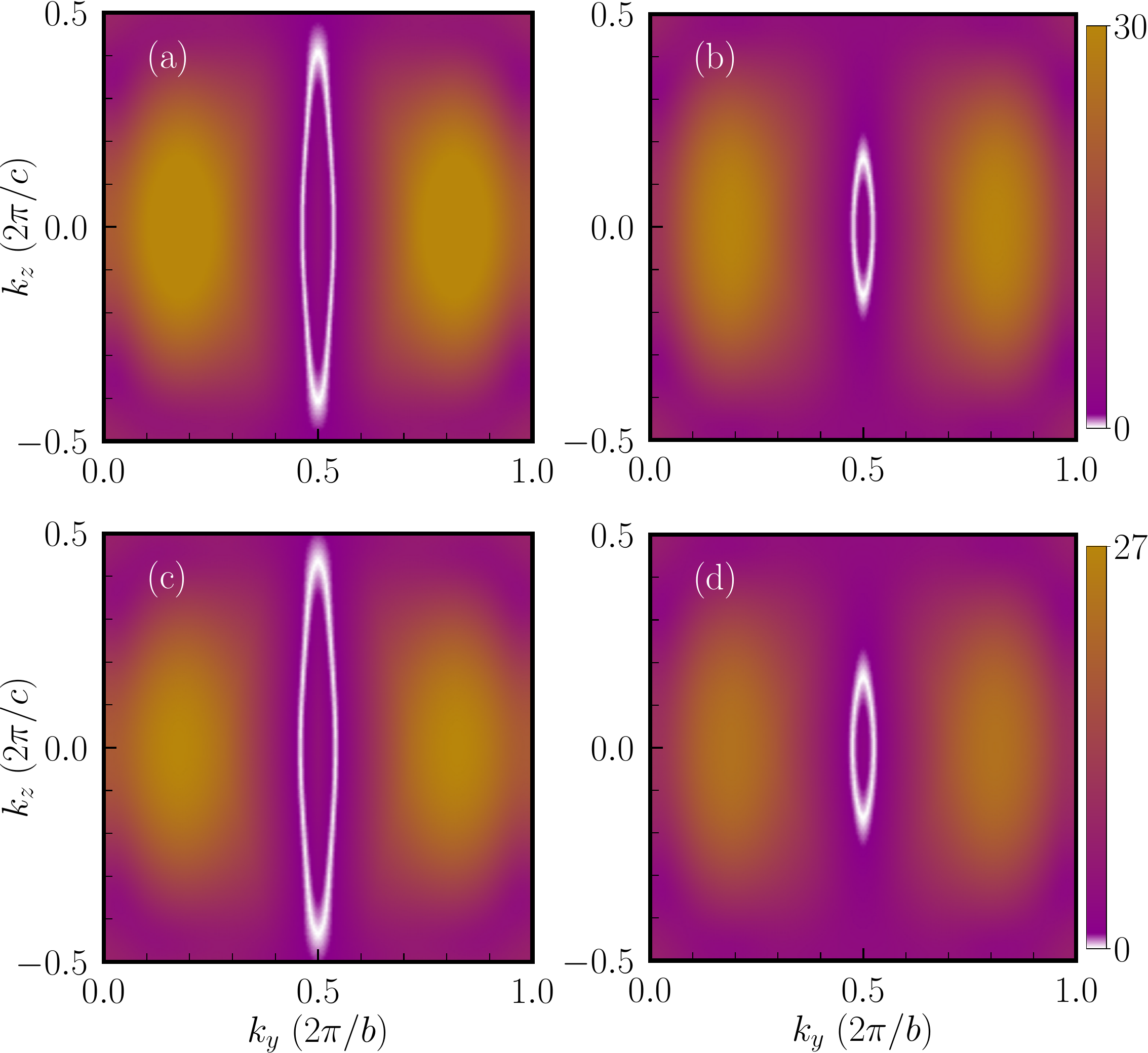}
\caption {\label{fig:deform} (a,b) Distribution of the Dirac ring (white) for TaOsSi and NbOsSi, encircling the $S$ point. (c,d) Dirac nodal rings for TaOsSi and NbOsSi with lattice parameters uniformly increased by $5$ \%. Notably, the Dirac ring remains intact under lattice distortions that preserve the underlying crystal symmetry. The color map indicates the magnitude of the local band gap in meV.}
\end{figure}

Upon incorporating SOC, TaOsSi and NbOsSi host a Dirac nodal ring encircling the $S$ point within the $k_x=\pi$ plane, protected by the glide-mirror symmetry $\mathcal{G}_x$, as shown in Figure~\ref{fig:deform}(a) and \ref{fig:deform}(b). To further assess the robustness of this Dirac nodal structure, we introduce a uniform lattice deformation of up to $5\%$ in both TaOsSi and NbOsSi. As shown in Figure~\ref{fig:deform}(c) and \ref{fig:deform}(d), the Dirac ring remains intact under such deformation, exhibiting neither gap opening nor any qualitative change in its topology. This resilience directly reflects the protection provided by the glide-mirror symmetry $\mathcal{G}_x$, which is preserved under the applied lattice distortion. Consequently, the symmetry-enforced band connectivity responsible for the Dirac nodal ring is maintained, underscoring the robustness of this topological feature against realistic structural perturbations.

\section*{Minimal model for hourglass dispersion and topological superconductivity}
We consider a semi-infinite geometry occupying the half-space $y\leq 0$, with the surface normal to the $y$-direction. For this geometry, the in-plane momenta $k_z$ remain good quantum numbers, while $k_y$ is replaced by the operator $-i\partial_y$. To implement the open boundary condition, we discretize the $y$ coordinate and recast the system as a stack of coupled quintuple layers with a surface located at $y=0$. Each quintuple layer is labeled by an integer index $n$, and the spatial derivatives are approximated as $\partial_y\psi_n(z)=\frac{1}{2}\left[\psi_{n+1}-\psi_{n-1}\right]$ and $\partial_{y}^2 \psi_n(z)=\left[ \psi_{n+1}+\psi_{n-1}-2\psi_n \right]$, where the lattice constant along the $z$-direction is set to unity.

Within this discretized representation, the Hamiltonian naturally separates into intralayer and interlayer contributions, $H_{\mathrm{BdG}}=H_\parallel + H_\perp$. The intralayer term is given by $H_\parallel=\sum_{n,k_z}\psi_{n,k_z}h_\parallel(k_z)\psi_{n,k_z}$, with

\begin{align}
h_\parallel(k_z) &= t~ \kappa_z \sigma_x +  t_{\rm c}~\kappa_z \tau_x + t_a ~\kappa_z \sigma_x \tau_z + \lambda~\kappa_z s_y \sigma_z \tau_y + t_\perp (\cos k_z \kappa_z \tau_x - \sin k_z \kappa_z \tau_y) \nonumber \\
&+ \lambda_\perp (\cos k_z \kappa_z s_y \sigma_z \tau_y + \sin k_z \kappa_z s_y \sigma_z \tau_x) + \frac{1}{2} (m_1 + m_2) + \frac{1}{2} (m_1 - m_2) \kappa_z \sigma_z \tau_z \nonumber\\
&+ \lambda_{\perp}^\prime \sin k_z \kappa_z s_x \sigma_z \tau_z\,.
\end{align}

\noindent Here, $t$ sets the energy scale of the hourglass dispersion, while $t_{\rm c}$ hybridizes the chain degrees of freedom. The parameter $t_a$ controls the symmetry-enforced partner switching responsible for the hourglass neck, and $\lambda$ denotes the nonsymmorphic spin–orbit coupling that protects the crossings. Interladder hopping and its associated spin–orbit coupling are described by $t_\perp$ and $\lambda_\perp$, respectively. The parameters $m_1$ and $m_2$ tune the position of the Dirac crossings relative to the Fermi level, while $\lambda_{\perp}^\prime$ lifts residual degeneracies. 

The interlayer hopping is described by $H_\perp=\sum_{n,k_z}\psi_{n,k_z}h_\perp\psi_{n+1,k_z} + \text{H.c.}$, where

\begin{equation}
	h_\perp= \frac{t}{2} \bigg( \kappa_z \sigma_x -i \kappa_z \sigma_y \bigg) + \frac{t_a}{2} \bigg( \kappa_z \sigma_x \tau_z -i \kappa_z \sigma_y \tau_z \bigg).
\end{equation}

\noindent To incorporate superconductivity, we introduce the INT pairing and define the Nambu spinor $\Phi_{n,k_z}=\left[ \psi_{n,k_z}^\dagger, \psi_{n,-k_z} \right]$. The intralayer Bogoliubov–de Gennes (BdG) Hamiltonian then takes the form $H_{\parallel}^{SC}=\sum_{n,k_z}\Phi_{n,k_z}^\dagger H_{SC}(k_z)\Phi_{n,k_z}$, where
\begin{equation}
	H_{SC}(k_z)=
	\begin{bmatrix}
		h_0(k_z) & \Delta \\
		-\Delta^* & -h_{0}^*(-k_z)
	\end{bmatrix},
\end{equation}
\noindent where $h_0(k_z)=h_\parallel(k_z)-\mu I_{16\times 16}$. Neglecting interlayer pairing, the interlayer contribution to the BdG Hamiltonian becomes $H_{\perp}^{SC}=\sum_{n,k_z}\Phi_{n,k_z}^\dagger H_{\perp}\Phi_{n+1,k_z}$, with 
\begin{equation}
	H_\perp=
	\begin{bmatrix}
		h_\perp & 0 \\
		0 & - h_{\perp}^*
	\end{bmatrix}.
\end{equation}
\noindent The retarded surface Green’s function $G(\bar{k},\omega)$ is computed using the standard transfer-matrix formalism~\cite{Wang2010,Sancho_1984}. Here, $\bar{k}$ denotes the surface momentum along the $\hat{z}$ direction in the surface Brillouin zone, while $\omega$ represents the energy of the surface states. It satisfies
\begin{subequations}
	\begin{align}
		G^{-1} &=g^{-1}-H_{\perp}^\dagger T, \\
		T &= GH_{\perp},
	\end{align}
\end{subequations}
\noindent where $g=[(\omega + i\eta)I_{32\times 32}-H_{SC}(k_z)]^{-1}$ is the Green’s function of an isolated layer and $\eta$ is a positive infinitesimal. The calculation is initialized with $G=g$ and iterated to self-consistency. Once the surface Green’s function is obtained, the surface spectral function is evaluated as
\begin{equation}
	A(\bar{k},\omega)=-\frac{1}{\pi}Im\left[ \operatorname{Tr}G(\bar{k},\omega) \right].
\end{equation}

\end{document}